\documentclass[twocolumn]{article}
\usepackage[margin=2cm]{geometry}
\usepackage{soul}
\usepackage[title]{appendix}
\usepackage{xcolor}
\usepackage{float}
\usepackage{graphicx} 
\usepackage{hyperref}
\usepackage{amsmath}
\usepackage{booktabs} 
\usepackage{ragged2e}
\usepackage{subcaption}
\usepackage[percent]{overpic}
\usepackage{xcolor}
\usepackage{placeins}
\usepackage{makecell}
\usepackage[utf8]{inputenc}
\usepackage[T1]{fontenc}
\soulregister\cite7
\usepackage[
  backend=biber,
  style=numeric-comp,
  sorting=none,
  maxbibnames=99,
  maxcitenames=2
]{biblatex}

\AtEveryBibitem{%
  \clearfield{urldate}%
  \clearfield{url}%
  \clearfield{urlyear}%
  \clearfield{urlmonth}%
  \clearfield{urlday}%
  \clearfield{note}%
  \clearfield{eprint}%
}
\title{Coupling Heterarchical Granular Dynamics and Computational Fluid Dynamics}
\author{Jiahuan Li, Shivakumar Athani, Alistair Gillespie, Matthew Cleary, Itai Einav and Benjy Marks}
\date{}

\begin{document}

\justifying      
\setlength{\parindent}{10pt} 

\maketitle
\begin{abstract}
Granular flows in ambient fluids exhibit grain-size-dependent segregation, which is difficult to capture efficiently with existing models, especially in large-scale systems involving more than a million grains. We develop a two-way coupled framework that integrates heterarchical granular dynamics (HGD) with a fluid-fraction-weighted incompressible Navier–Stokes solver. This heterarchical granular-fluid dynamics (HGFD) model extends a previous HGD model for quasi-static deformations by introducing inertial, force-balance-driven particle velocities and consistent fluid-solid momentum exchange. The coupling between the inertial HGD and the fluid solver is performed using a staggered explicit sequential scheme and co-located Eulerian fields. The framework is evaluated against experimental data of (i) single-particle settling to verify inertial relaxation, (ii) hindered settling to reproduce concentration-dependent settling and vertical size stratification, and (iii) representative cases covering three reported segregation types to assess regime sensitivity. These results establish HGFD as an efficient and consistent approach for simulating fluid-coupled granular segregation dynamics.
\end{abstract}
\section{Introduction}

Many natural and industrial processes involve the interaction of granular materials and fluids, including submarine landslides, mud and slurry flows, fluidised-bed operations, and  various other systems in mineral processing \cite{stickel_fluid_2005,lee_twophase_2018,crowe_multiphase_2011,kunii_fluidization_2013,iverson_physics_1997}.
In all of these applications, the interaction between particle inertia, hydrodynamic drag, and concentration-dependent effects give rise to collective phenomena such as shear-induced migration, concentration stratification, and segregation by particle size and density \cite{richardson_sedimentation_1954,guazzelli_rheology_2018}. 
Accurate numerical prediction of such multi-phase interactions remains a significant challenge due to the wide range of spatial and temporal scales involved \cite{vanderhoef_numerical_2008}.

Existing models of particle-fluid systems can broadly be classified into continuum two-fluid models (TFM) and particle-resolved methods, including discrete element method (DEM) coupled with computational fluid dynamics (CFD), i.e., CFD--DEM \cite{anderson_fluid_1967, zhu_discrete_2008}. 
TFM treats both phases as interpenetrating media and represents particle stresses through constitutive relations and interphase momentum transfer through drag correlations \cite{anderson_fluid_1967, lun_kinetic_1984}. 
While computationally efficient, TFM does not retain particle-scale transport mechanisms explicitly and often requires empirical calibration for heterogeneous structures and segregation/mixing \cite{jackson_dynamics_2000,agrawal_role_2001}. 
In contrast, the CFD--DEM approach resolves particle motion individually and computes hydrodynamic forces at the grain level, providing detailed physical representation but at significant computational cost that limits its applicability to large-scale systems \cite{zhu_discrete_2008, jajcevic_largescale_2013}.

Heterarchical Granular Dynamics (HGD), originally introduced by Marks et al.~\cite{marks_heterarchical_2025}, provides an alternative mesoscopic description for particle transport. The use of the term heterarchy refers to the cross-scale passage of information: here, grain and pore properties are represented along an internal microstructural coordinate at each location in physical space. In HGD,  information exchanges probabilistically, freely through the composed physical-microstructural space $via$ advection and diffusion of voids. This formulation preserves essential transport mechanisms while maintaining computational efficiency through a cell-based stochastic representation.

However, the formulation of the original HGD framework was restricted for simplicity to quasi-static particle transport. This involved simplified prescribed kinematic rules that did not explicitly allow for inertial dynamics. Furthermore, the original formulation did not consider the possible interactions between particles and fluid \cite{marks_heterarchical_2025}. 
As a result, this previous formulation of HGD cannot be used for fluid-driven granular systems, particularly those involving momentum exchange and concentration-dependent drag \cite{richardson_sedimentation_1954,guazzelli_rheology_2018}.

The present work involves three major extensions. First, the HGD framework is enhanced by introducing explicit inertial particle dynamics governed by local force balance. Second, the extended granular phase is coupled to a continuum fluid solver, enabling two-way momentum exchange while retaining the mesoscopic stochastic transport structure of HGD.
Third, the integration of HGD is realised through a staggered explicit sequential scheme with cell-wise exchange of solid fraction, solid-phase velocity, and an interphase momentum source term in the fluid momentum equation.

The objectives of this study are therefore threefold: 
(i) to formulate a consistent inertial extension of HGD compatible with fluid--particle force balance; 
(ii) to develop a numerically stable two-way coupling strategy with a fluid-fraction-weighted incompressible Navier-Stokes solver; and 
(iii) to demonstrate that the resulting framework captures both 
particle-level relaxation dynamics and collective segregation phenomena.

The present work is restricted to two-dimensional laminar flow 
conditions; extension to turbulent regimes is left for future study.

\section{Methodology}

\subsection{Heterarchical granular dynamics}

\label{sec:heterarchy}
Heterarchy is an organisational principle in which components at different scales are not separated into distinct levels but instead coexist and exchange information within a single framework. In the context of granular materials, this concept was introduced by Marks and Einav \cite{marks2017heterarchical} as a multiscale modelling paradigm in which the microstructure is represented through an internal coordinate independent of the spatial and temporal coordinates. This concept has since been applied to problems including comminution in rotary mills \cite{bisht_heterarchical_2024,bisht_heterarchical_2024a,bisht_heterarchical_2025} and granular flow dynamics \cite{marks_heterarchical_2025}. The present work builds on the Heterarchical Granular Dynamics (HGD) framework introduced by Marks et al.~\cite{marks_heterarchical_2025}, which describes granular transport at the mesoscopic scale.

The HGD framework is formulated on a heterarchical lattice defined over physical spatial dimensions $(x,y)$ augmented by an internal microstructural coordinate $m$. For each spatial position on the lattice $(i,j)$, the microstructural coordinate is discretised into $M$ internal coordinates indexed by $k = 1,\dots,M$, each occupied either by a grain or by a void. The ensemble of internal coordinates within a given spatial position constitutes a representative volume element (RVE) of the granular assembly at that location. Before presenting the transport rules, we summarise the key quantities used in the formulation that follows. The number of solid-occupied coordinates is $M_s$ and the number of void-occupied coordinates is $N = M - M_s$, giving the local solid volume fraction $\phi = M_s/M$ and the void fraction $n = N/M = 1 - \phi$. Each solid-occupied coordinate $k$ is assigned a particle size $s_k$. The lattice spacings are $\Delta x$ and $\Delta y$ in the horizontal and vertical directions respectively, and $\Delta t$ is the discrete time step. The critical solid fraction $\phi_c$ denotes the jammed state corresponding to the maximum packing of the solid phase, and $\alpha$ is the mixing coefficient governing the diffusivity $D$.

Within this heterarchical lattice, particle transport is represented through stochastic migration of voids between neighbouring cells. At each discrete time step, voids are exchanged according to probabilistic advection and diffusion rules, which collectively represent particle motion when averaged over the heterarchical coordinate. Since each exchange swaps a void with a neighbouring solid-occupied coordinate, it can be described equivalently from the perspective of either the void or the solid element, with the two representations kinematically identical and conserving the local number of grains and voids by construction. We retain the void-based description of Marks et al.~\cite{marks_heterarchical_2025} for consistency with the original HGD formulation.

\subsubsection{Original HGD formulation} 

In the original HGD model~\cite{marks_heterarchical_2025}, gravity is aligned in the $-y$ direction, such that voids advect upward (in $+y$) by swapping with neighbouring cells, while particles move downward. In this model, the characteristic velocity of void migration is gravity-driven: neglecting inertia, a particle is assumed to traverse one cell spacing under gravity before coming to rest. The resulting velocity scale is
\begin{equation}
u = \sqrt{g\,\Delta y},\label{eq:u_qs}
\end{equation}
where $g$ is the gravitational acceleration. The void advection is implemented stochastically by moving a distance $\Delta y$ within a time step $\Delta t$ with probability $P_{\mathrm{adv}}$, such that the mean velocity satisfies $u = P_{\mathrm{adv}}\Delta y/\Delta t$. This gives
\begin{equation}
P_{\mathrm{adv}} = u\,\frac{\Delta t}{\Delta y}.\label{eq:Padv_basic}
\end{equation}

In addition to vertical advection, voids diffuse in the $\pm x$ direction by random motion, with a diffusivity $D$ leading to a diffusion probability per time step
\begin{equation}
P_{\mathrm{diff}} = D\,\frac{\Delta t}{\Delta x^{2}}.
\label{eq:Pdiff}
\end{equation}
The original HGD model further assumes that the diffusivity is proportional to the advection velocity and the local mean particle size, with constant of proportionality $\alpha$, i.e.
\begin{equation}
\alpha = \frac{D}{u\,s_k},
\label{eq:alpha_basic}
\end{equation}
so that, equivalently, $D = \alpha\,u\,s_k$, with $s_k$ evaluated at the destination cell into which the void diffuses.
\begin{figure*}[!t]
\centering
\includegraphics[width=\textwidth]{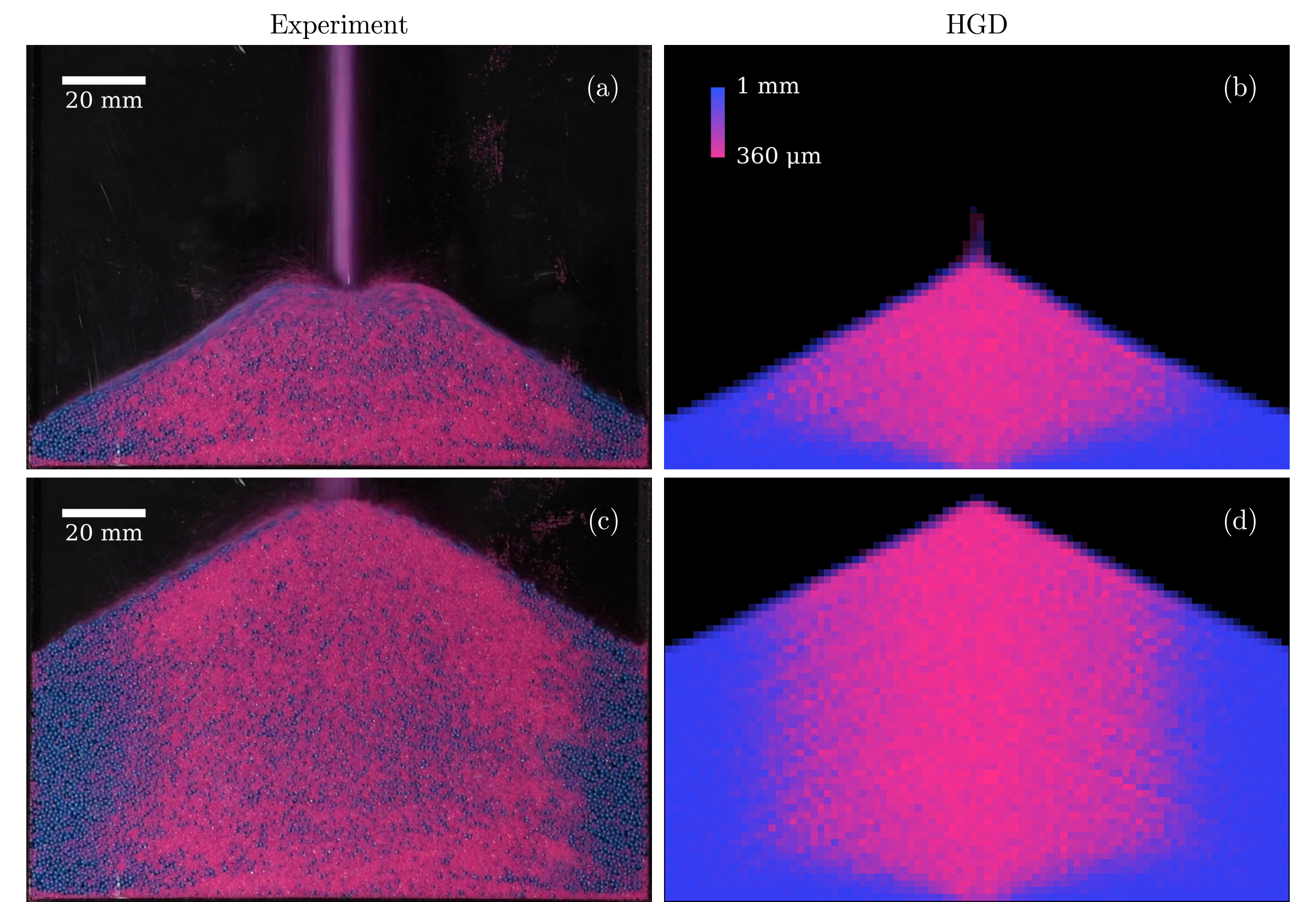}
\caption{
Comparison of bidisperse granular filling (1\,mm and 360\,\textmu m) between experimental observations (a, c) and the corresponding quasi-static HGD simulations without fluid coupling (b, d). Panels (a, b) show an early stage and panels (c, d) a later stage.
}
\label{fig:motivation_patterns}
\end{figure*}

\subsubsection{Motivation}
\label{sec:Motivation}

The original HGD formulation described above is restricted to quasi-static granular rearrangements in which particles remain in sustained contact and the influence of any surrounding fluid is neglected. Within this regime, the quasi-static HGD framework is nevertheless sufficient to reproduce the main deposition and segregation features observed experimentally during granular filling (Fig.~\ref{fig:motivation_patterns}). 

When particles gain non-negligible inertia in dry systems, bulk deformation can no longer be described quasi-statically. When immersed in a fluid, particles are influenced by hydrodynamic forces such as drag and buoyancy, and the particle velocity is determined by a local force balance. These mechanisms introduce velocity relaxation towards the surrounding fluid motion and lead to momentum exchange between the granular and fluid phases. As a result, the particle transport cannot be fully captured by the quasi-static kinematic description of the original HGD model.

\begin{figure*}[t!]
    \centering
    \includegraphics[width=\textwidth]{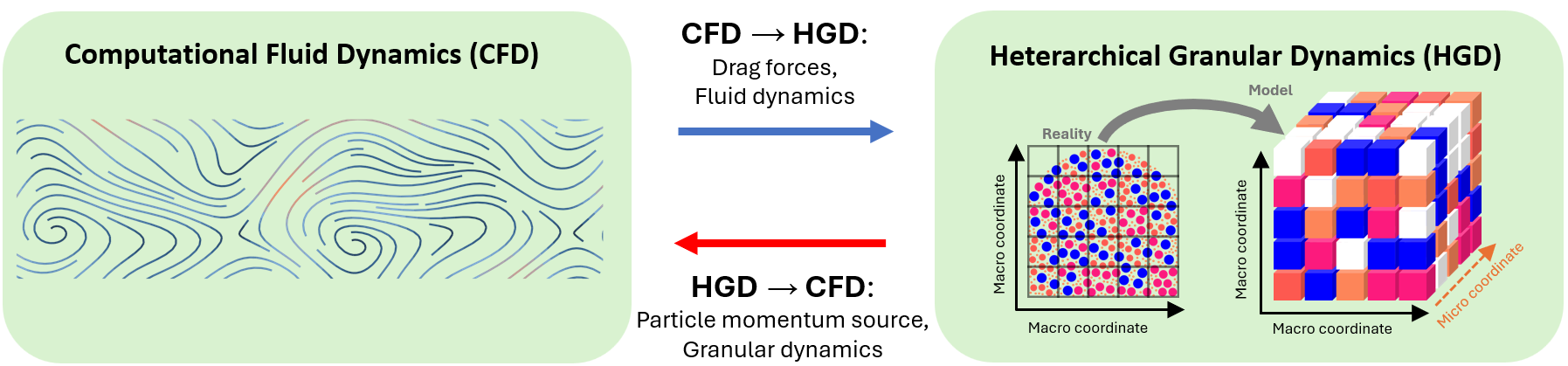}
    \caption{Schematic representation of the two-way coupling framework.}
    \label{fig:data_transfer}
\end{figure*}

To address this limitation, the present work extends the HGD formulation to inertial, fluid-coupled granular systems, as illustrated in Figure~\ref{fig:data_transfer}. Because HGD resolves the particle size distribution through internal coordinates at each spatial position, the extension introduces a velocity and a force balance at each internal coordinate independently, and couples the drag computed at each internal coordinate to a continuum fluid solver through aggregated momentum exchange. This construction allows size-dependent drag and segregation to emerge directly from the heterarchical description without requiring separate transport equations for each particle size class.

\subsubsection{HGD for fluid-coupled, inertial dynamics}
\label{sec:fluid-coupled-hgd}

In contrast to the original HGD formulation~\cite{marks_heterarchical_2025},  where particle motion is characterised by a prescribed scalar velocity stated in Eq.~\ref{eq:u_qs}, the extended model introduces a particle velocity vector $\vec{u}_k = (u_{x,k}, u_{y,k})$, defined at each spatial position $(i,j)$ and internal coordinate $k$, that evolves dynamically in time. The particle  velocity is updated explicitly at each time step according to
\begin{equation}
    \vec{u}_k^{\,t+\Delta t} = \vec{u}_k^{\,t} 
    + \vec{a}_k^{\,t}\,\Delta t,
\end{equation}
where $\vec{a}_k^{\,t}$ is the net acceleration at internal coordinate $k$, comprising buoyancy-modified gravity and fluid drag,
\begin{equation}
    \vec{a}_k = \vec{a}_{\mathrm{b}} 
    + \vec{a}_{\mathrm{d,k}}.
\label{eq:particle_accel}
\end{equation}
The two contributions are defined below.

\paragraph{Buoyancy-modified gravitational acceleration.}
The gravitational contribution is modified by buoyancy to account for the reduced weight of a particle immersed in fluid. By Archimedes' principle, the net gravitational force on a submerged particle is the difference between its weight and the buoyancy force, $\vec{F} = (\rho_p - \rho_f)V\vec{g}$~\cite{gidaspow_multiphase_1994}. Dividing by the particle mass $\rho_p V$ yields the effective gravitational acceleration,
\begin{equation}
    \vec{a}_{\mathrm{b}}=\left(1-\frac{\rho_f}{\rho_p}\right)\vec{g},
\label{eq:ab}
\end{equation}
where $\vec{g}$ is the gravitational acceleration vector, and $\rho_f$ and $\rho_p$ denote the fluid and particle densities, respectively.

\paragraph{Drag-induced acceleration.}

When a particle moves relative to the surrounding fluid, the fluid exerts a resistive drag force proportional to the velocity difference $(\vec{u}_f - \vec{u}_k)$~\cite{anderson_fluid_1967}, where $\vec{u}_f$ is the local fluid velocity. This force drives the particle velocity towards the local fluid velocity over a characteristic relaxation time $\tau_{p,k}$,
\begin{equation}
    \vec{a}_{\mathrm{d},k} = \frac{1}{\tau_{p,k}}(\vec{u}_f - \vec{u}_k), \qquad \tau_{p,k} = \frac{\rho_p}{\beta_k},
\label{eq:ad}
\end{equation}
where $\beta_k$ is the momentum-exchange coefficient at internal coordinate $k$, defined below.

\paragraph{Interphase momentum-exchange coefficient.}
The momentum-exchange coefficient $\beta_k$ at each solid-occupied internal coordinate $k$ depends on the local solid volume fraction $\phi$:

\begin{equation}
\beta_k=
\begin{cases}
\beta_{k,\mathrm{WY}}, & \phi < 0.2,\\[4pt]
\beta_{k,\mathrm{Er}}, & \phi \ge 0.2.
\end{cases}
\label{eq:beta_switch}
\end{equation}
This piecewise form follows the combined drag model of Gidaspow~\cite{gidaspow_multiphase_1994}, which is widely adopted in Eulerian two-fluid models to cover the full range of solid concentrations encountered in fluid-coupled granular systems. The two limbs are chosen because no single correlation is accurate across all concentrations: the Wen--Yu coefficient $\beta_{k,\mathrm{WY}}$~\cite{wen_mechanics_1966} is derived from the single-particle drag law with a voidage correction, and is therefore appropriate in the dilute-to-moderate regime where particles interact primarily through the surrounding fluid; the Ergun-type coefficient $\beta_{k,\mathrm{Er}}$~\cite{ergun_fluid_1952} is derived from packed-bed pressure-drop measurements, treating the dense assembly as flow through a network of tortuous pore channels, and is therefore appropriate once particles are in near-permanent proximity. The threshold $\phi = 0.2$ marks the transition between these two physical pictures and is the value commonly used in two-fluid models~\cite{gidaspow_multiphase_1994}, ensuring continuity of the drag force across the switch.

The Wen--Yu coefficient at coordinate $k$ is
\begin{equation}
\beta_{k,\mathrm{WY}}= \frac{1}{M_s}\,
0.75\,C_{d,k}\,\rho_f\,\frac{|\vec{u}_f-\vec{u}_k|}{s_k}\,\phi\,n^{-2.65},
\label{eq:beta_wy}
\end{equation}

where $\rho_f$ is the fluid density and the $-2.65$ exponent is a Richardson--Zaki-type hindered-settling correction.

The drag coefficient $C_{d,k}$ incorporates inertial corrections through the Schiller--Naumann correlation~\cite{schiller_uber_1933}:
\begin{equation}
C_{d,k}=\frac{24}{Re_{p,k}}\left(1+0.15\,Re_{p,k}^{0.687}\right),
\label{eq:Cd}
\end{equation}

where the particle Reynolds number at internal coordinate $k$ is
\begin{equation}
Re_{p,k}=\frac{\rho_f\,|\vec{u}_f-\vec{u}_k|\,s_k}{\mu_f},
\label{eq:Rep}
\end{equation}
and $\mu_f$ is the fluid dynamic viscosity.

For dense conditions, the Ergun-type coefficient at coordinate $k$ is
\begin{equation}
\beta_{k,\mathrm{Er}}=\frac{1}{M_s}\left(
\frac{150\,\mu_f\,\phi^2}{n\,s_k^2}
+
1.75\,\rho_f\,\phi\,\frac{|\vec{u}_f-\vec{u}_k|}{s_k}\right).
\label{eq:beta_ergun}
\end{equation}

The expressions above define $\beta_k$ at each internal coordinate $k$. For the two-way coupling with the fluid solver (Section~\ref{sec:two_way_coupling}), a cell-level momentum-exchange is required and is obtained by summing over all internal coordinates at a given spatial position, $\beta = \sum_k \beta_k$, where $\beta_k = 0$ for coordinates not occupied by solid.


\paragraph{Exponential time integration.}
Substituting the buoyancy (Eq.~\ref{eq:ab}) and drag (Eq.~\ref{eq:ad}) terms into Eq.~\ref{eq:particle_accel}, we get a differential equation for the particle velocity. Assuming constant coefficients at any given time step, the
equation for the particle velocity becomes ordinary, which could be solved analytically to give:
\begin{equation}
\vec{u}_k^{\,t+\Delta t} = \vec{u}_f + \tau_{p,k}\,\vec{a}_{\mathrm{b}} 
+ \left(\vec{u}_k^{\,t} - \vec{u}_f - \tau_{p,k}\,\vec{a}_{\mathrm{b}}\right)
\exp(-\Delta t/\tau_{p,k}).
\label{eq:exact}
\end{equation}
A similar approach is widely adopted in Euler–Lagrange simulations of dispersed multiphase flows \cite{shirolkar_fundamental_1996,goz_study_2004}. It was shown that this method avoids the severe time-step restrictions imposed by explicit schemes in stiff drag regimes.

\paragraph{Directional advection.}
The updated particle velocity $\vec{u}_k$ sets the direction of the void--solid swaps introduced in Section~\ref{sec:heterarchy}, and thereby the advection of voids between neighbouring cells. In the original HGD model, advection is restricted to the vertical direction because particle motion is driven solely by gravity. In the fluid-coupled formulation, however, drag and buoyancy forces can drive particle motion in any direction. To capture this, the advection probability is generalised from the vertical-only form (Eq.~\ref{eq:Padv_basic}) to act independently in each spatial direction using an upwind decomposition:
\begin{equation}
P_{\xi,+,k} = \frac{(u_{\xi,k})^{+} \Delta t}{\Delta x}, \qquad
P_{\xi,-,k} = \frac{(u_{\xi,k})^{-} \Delta t}{\Delta x},
\end{equation}
where $(u)^{+} = \max(u,0)$ and $(u)^{-} = \max(-u,0)$, $u_{\xi,k}$ is the velocity component in direction $\xi$ at internal coordinate $k$, and $\Delta x$ is the spatial grid size. The direction of advection is determined by the sign of the velocity, and local mass conservation follows from the swap mechanism described in Section~\ref{sec:heterarchy}.

\paragraph{Concentration-dependent diffusion.}
In the original HGD formulation, the diffusivity is proportional to the advection velocity and the local particle size with a constant coefficient $\alpha$ (Eq.~\ref{eq:alpha_basic}). This constant treatment is appropriate for dense flows where the solid fraction remains near the critical value $\phi_c$. In fluid-coupled systems, however, the solid fraction can vary from nearly dilute to densely packed, and the intensity of diffusive mixing is expected to increase with solid fraction as inter-particle interactions become more frequent. To capture this behaviour, we generalise $\alpha$ to depend on the local solid fraction,
\begin{equation}
D_k = \alpha(\phi)\,s_k\,|u_{y,k}|.
\label{eq:diffusivity}
\end{equation}

\begin{figure}[!t]
    \centering
    \includegraphics[width=\columnwidth]{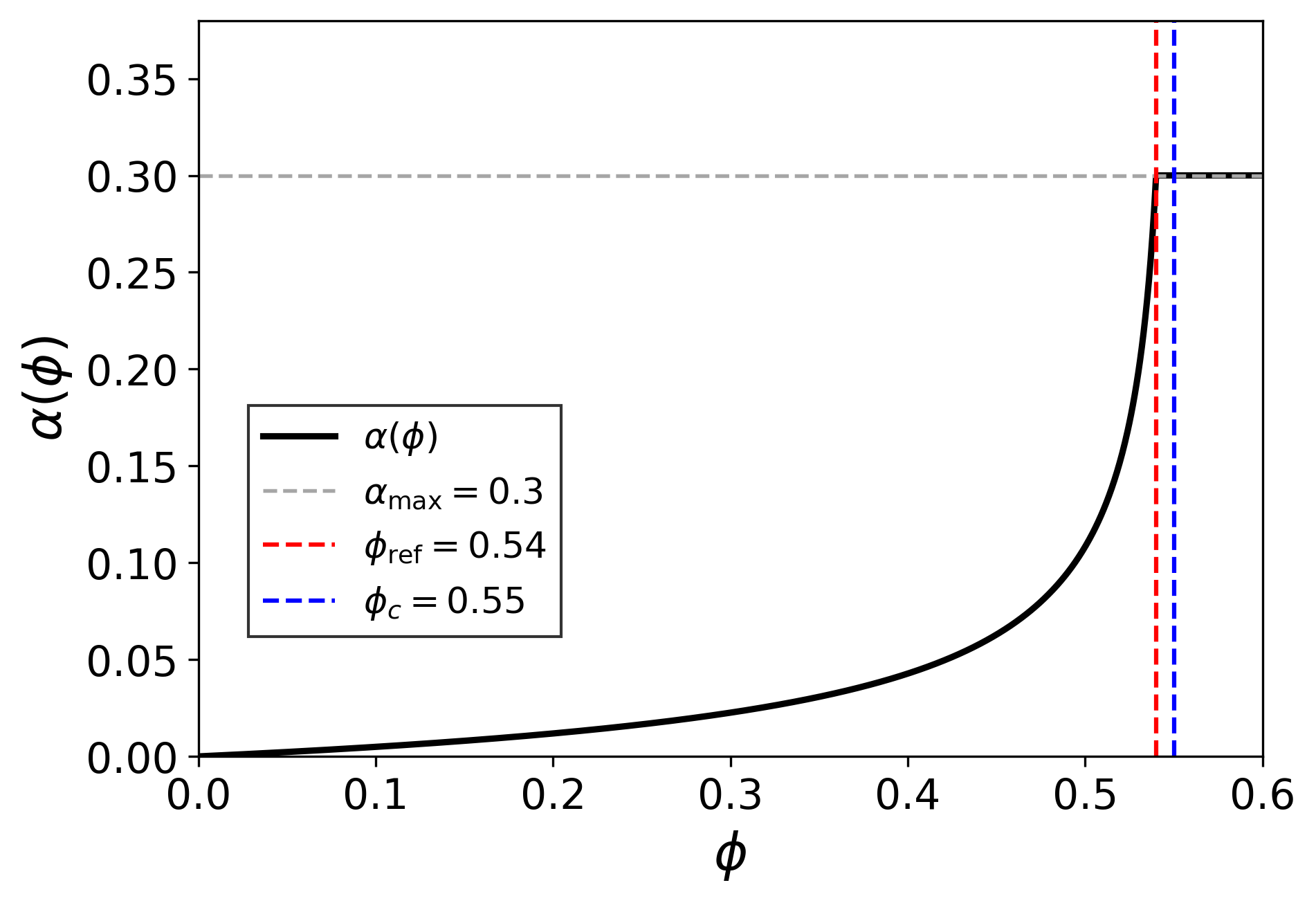}
    \caption{Concentration-dependent mixing coefficient $\alpha(\phi)$ defined by Eq.~\eqref{eq:alpha_map}, with $\phi_c = 0.55$, $\phi_{\mathrm{ref}} = 0.54$, and $\alpha_{\max} = 0.3$.}
    \label{fig:alpha_phi}
\end{figure}

Following \cite{athani_unifying_2024}, who showed that the normalised shear-induced self-diffusivity scales as $(\phi_c - \phi)^{-1/2}$ across both suspension and granular regimes, we propose the following construction for $\alpha(\phi)$, designed to vanish in the dilute limit and saturate at the dense quasi-static value $\alpha_{\max}$:
\begin{equation}
\alpha(\phi) = \min\!\left(\alpha_{\max},\; \alpha_{\max}\, \frac{(\phi_c - \phi)^{-1/2} - \phi_c^{-1/2}}{(\phi_c - \phi_{\mathrm{ref}})^{-1/2} - \phi_c^{-1/2}}\right)
\label{eq:alpha_map}
\end{equation}

where $\phi_{\mathrm{ref}} = \phi_c - \epsilon$ with $\epsilon = 0.01$ is introduced to prevent divergence at $\phi = \phi_c$, and $\alpha_{\max} = 0.3$. An illustrative profile using the parameters of Case~3 (Table~\ref{tab:sim_params}, $\phi_c = 0.55$) is shown in Figure~\ref{fig:alpha_phi}. Note that in the current version of the HGD framework the solid fraction is bounded by $\phi \leq \phi_c$ by construction, since the critical solid fraction defines the maximum packing state of the heterarchical lattice~\cite{marks_heterarchical_2025}. Only the $(\phi_c - \phi)^{-1/2}$ functional form is adopted from \cite{athani_unifying_2024}; the prefactor $\alpha_{\max}$ is independently calibrated, see Section~\ref{sec:seg_types}. The remaining parameter values are listed in Table~\ref{tab:sim_params}. The resulting diffusivity is used within the standard HGD diffusion probability (Eq.~\ref{eq:Pdiff}).

\subsection{Fluid Governing Equations}
Having defined the heterarchical transport physics of the granular phase, we now specify the fluid-phase description and the coupling between the two phases. Because the solid and fluid phases coexist within each computational cell, the standard incompressible Navier--Stokes equations are replaced by their fluid-fraction-weighted counterparts~\cite{anderson_fluid_1967}, in which the fluid volume fraction $n$ represents the space available to the fluid phase. Assuming incompressible flow and rigid particles with no phase change, the continuity and momentum equations for the fluid phase read

\begin{equation}
\nabla \cdot (n \vec{u}_f) = 0,
\label{eq:mass_conservation_void}
\end{equation}
and
\begin{equation}
\frac{\partial(n\vec{u}_f)}{\partial t}
+ \nabla \cdot (n\vec{u}_f \otimes \vec{u}_f)
= -\frac{n}{\rho_f}\nabla p
+ \nabla \cdot (n\boldsymbol{\tau})
+ \frac{\vec{S}_m}{\rho_f},
\label{eq:momentum_conservation}
\end{equation}

where $\vec{u}_f$ is the fluid velocity, $p$ is the fluid pressure, $\rho_f$ is the fluid density, $\boldsymbol{\tau} = \nu (\nabla \vec{u}_f + \nabla \vec{u}_f^T)$ is the kinematic viscous stress tensor with constant kinematic viscosity $\nu = \mu/\rho_f$, and $\vec{S}_m = \beta(\vec{u}_f - \vec{u})$ is the interphase momentum source term per unit volume, evaluated at each spatial position $(i,j)$ from the drag force computed by the HGD model, with $\beta$ and $\vec{u}$ defined in Eq.~\eqref{eq:aggregation}.


\subsection{Mass and momentum conservation}
While the HGD computation is performed stochastically on the heterarchical lattice, it is useful to examine the continuum limit of the formulation to verify consistency with established conservation laws. Both the mass and momentum conservation equations are obtained by extending the derivation in~\cite{marks_heterarchical_2025} to include directional advection and concentration-dependent diffusion, with full details given in Appendix~\ref{app:mass} and Appendix~\ref{app:momentum}.

The continuum form of the mass conservation equation for the void fraction is obtained by taking the lattice-based discrete formulation to the limit as $\Delta x$, $\Delta y$, and $\Delta t$ approach zero:
\begin{equation}
\begin{split}
\frac{\partial n}{\partial t} 
&= \frac{\partial}{\partial x}\!\left[
    D\,\frac{\partial n}{\partial x}
    + n\phi\,\frac{\partial D}{\partial x}
  \right] \\
&\quad - \frac{\partial}{\partial x}\left( u_x\, \phi\, n \right)
       - \frac{\partial}{\partial y}\left( u_y\, \phi\, n \right),
\end{split}
\label{eq:hgd_full_diff}
\end{equation}
where $u_x$ and $u_y$ are the components of the particle velocity averaged over the internal coordinates at each spatial position $(i,j)$, $D = \alpha(\phi)\,\overline{s_k\,|u_{y,k}|}$ is the diffusivity averaged over the internal coordinates, and $\alpha(\phi)$ is the concentration-dependent mixing coefficient defined in Eq.~\eqref{eq:alpha_map}. The first term on the right-hand side represents diffusive mixing, while the second and third terms represent advective transport in the $x$ and $y$ directions, modulated by the factor $n\phi$, which arises from the discrete swapping mechanism: a void migration can occur only when a void is available at the origin and solid material is available at the destination.

For momentum, since diffusive exchanges are symmetric, they do not contribute to net momentum transfer in the continuum limit, and only advective and external force terms remain:
\begin{align}
\frac{\partial(\phi\, u_x)}{\partial t} 
&= -\frac{\partial}{\partial x}(\phi\, u_x^2\, n)
   -\frac{\partial}{\partial y}(u_y\, \phi\, u_x\, n) + \phi F_x, \\[8pt]
\frac{\partial(\phi\, u_y)}{\partial t} 
&= -\frac{\partial}{\partial x}(u_x\, \phi\, u_y\, n)
   -\frac{\partial}{\partial y}(\phi\, u_y^2\, n) + \phi F_y,
\end{align}
where $\vec{F} = (F_x, F_y)$ denotes the net force per unit mass acting on the solid phase, including gravitational and drag contributions.

\subsection{Two-way coupling}
\label{sec:two_way_coupling}

To capture the bidirectional interaction between the fluid and granular phases, we develop a heterarchical granular--fluid dynamics (HGFD) two-way coupling solver. The overall coupling framework, illustrated in Figure~\ref{fig:data_transfer}, is advanced at each global time step using a staggered explicit partitioned coupling scheme\cite{preCICEv2}. The CFD and HGD modules are solved sequentially, exchanging data once per time step. Specifically, the HGD module is first advanced using the fluid velocity from the previous time step, yielding the updated granular state, including the local solid fraction, solid-phase velocity, and interphase momentum exchange quantities. These fields are then supplied to the CFD solver, which advances the fluid velocity and pressure fields.

\paragraph{CFD $\rightarrow$ HGD.} 
The CFD module provides the fluid velocity field $\vec{u}_f$ to the HGD model for evaluating the local phase-relative motion and drag-induced acceleration. In the present implementation, the CFD mesh and the HGD grid are co-located with a one-to-one correspondence, so the fluid velocity passed to HGD is taken directly as the cell-centred CFD velocity at each computational cell $(i,j)$.

\paragraph{HGD $\rightarrow$ CFD.} 
The HGD model resolves particle dynamics across multiple heterarchical layers $k$ within each cell $(i,j)$. For coupling with the CFD solver, these layer-wise quantities are aggregated into cell-level Eulerian fields:
\begin{equation}
\phi = \sum_k \phi_k, \quad
\vec{u} = \frac{\sum_k \beta_k \vec{u}_k}{\sum_k \beta_k}, \quad
\beta = \sum_k \beta_k,
\label{eq:aggregation}
\end{equation}

where the subscript $k$ denotes layer-wise quantities: $\phi_k$ is the solid fraction of layer $k$ (Section~\ref{sec:fluid-coupled-hgd}), $\vec{u}_k$ is the particle velocity at layer $k$ computed from Eq.~(\ref{eq:exact}), and $\beta_k$ is the corresponding momentum-exchange coefficient (Eq.~\ref{eq:beta_switch}). The solid-phase velocity $\vec{u}$ is defined as a $\beta_k$-weighted average so that the cell-level momentum exchange term $\beta(\vec{u}_f - \vec{u})$ is consistent with the total drag force exerted by all layers on the fluid. The aggregated fields assemble the interphase momentum source term $\vec{S}_m = \beta(\vec{u}_f - \vec{u})$ introduced in Eq.~\eqref{eq:momentum_conservation}, with the sign convention chosen such that the force exerted on the fluid is equal and opposite to the drag acting on the particles.

\paragraph{Implementation.} The CFD participant is implemented by extending the incompressible solver \texttt{pisoFoam} in OpenFOAM~\cite{weller_tensorial_1998}. At each coupling step, the solver reads the HGD-provided fields $\phi$, $\vec{u}$, and $\beta$, and incorporates the resulting interphase momentum exchange term into the fluid momentum equation. The data exchange and synchronisation between the CFD and HGD solvers are managed using the open-source coupling library preCICE~\cite{preCICEv2}.

\paragraph{Probability consistency and time-stepping.}
The HGD update is formulated in terms of probabilistic advection and diffusion events, with the event probabilities evaluated from the local state. To ensure probabilistic consistency and numerical robustness of the explicit update, the global time step $\Delta t$ is selected such that the total transition probability remains bounded throughout the domain:
\begin{equation}
P_{\mathrm{tot}} = \sum_{\xi} P_{\mathrm{adv}}^{\xi} + P_{\mathrm{diff}} < 0.5,
\end{equation}
where $P_{\mathrm{adv}}^{\xi}$ and $P_{\mathrm{diff}}$ denote the directional advection and diffusion event probabilities, respectively. The bound $P_{\mathrm{tot}} < 0.5$ ensures that at most one stochastic event is triggered per layer per time step, preventing competing transitions from occurring simultaneously and maintaining the stability of the staggered explicit HGFD coupling. A single global time step $\Delta t$ is shared by both the CFD and HGD modules, selected to satisfy both this constraint and the numerical stability requirements of the CFD solver. Unless otherwise stated, $\Delta t$ is kept constant within each simulation. The sensitivity of the solution to $\Delta t$, as well as to the spatial resolution and the number of heterarchical layers, is examined in Appendix~\ref{app:convergence}.

\paragraph{Momentum-transfer budget.}

The particle velocity $\vec{u}_k$ at each internal coordinate $k$, computed from the force balance (Eq.~\ref{eq:exact}), is a continuous quantity that evolves at every time step. However, particle migration in HGD is discrete: the solid element occupying a given coordinate can only change its spatial position when a stochastic swap event is triggered. As a result, a solid element may retain a non-zero velocity for several consecutive time steps without migrating to a neighbouring cell. If the drag source term is evaluated from this velocity throughout, momentum is repeatedly injected into the same CFD cell, leading to over-prediction of the interphase momentum transfer.

To prevent this artificial accumulation, we introduce a momentum-transfer budget that limits the total momentum a solid element can exchange with the fluid while residing in a given cell. For each internal coordinate $k$, a virtual displacement is accumulated from the particle velocity,
\begin{equation}
\delta x_{\mathrm{virt},k}^{t+\Delta t}
=
\delta x_{\mathrm{virt},k}^{t}
+
\left| \vec{u}_k^{\,t} \right| \Delta t.
\label{eq:virtual_disp}
\end{equation}
Once $\delta x_{\mathrm{virt},k} \ge \Delta x$, further drag-induced momentum transfer from that coordinate is suppressed until a discrete swap event occurs. When a swap takes place, the virtual displacement is reduced by one lattice spacing, $\delta x_{\mathrm{virt},k} \to \delta x_{\mathrm{virt},k} - \Delta x$, rather than reset to zero. This ensures that solid elements which migrate early carry a deficit into the next cell, while those that remain stationary beyond one cell-traversal time are prevented from injecting further momentum. In this way, the total momentum transferred per cell traversal remains consistent regardless of when the stochastic swap occurs.

Solid elements whose budget is exhausted are excluded from the $\beta_k$-weighted velocity average (Eq.~\ref{eq:aggregation}) when constructing the cell-level solid-phase velocity $\vec{u}$, while their momentum-exchange coefficients $\beta_k$ remain included in $\beta$. This reflects the fact that solid material continues to resist fluid motion regardless of whether a swap has occurred, whereas the rate of active momentum injection should cease once the budget at that coordinate is exhausted.

\subsection{Boundary conditions and spatial discretisation}
On the CFD side, no-slip conditions are applied at all solid walls, and the pressure is fixed at open boundaries. The grid resolution for each case is listed in Table~\ref{tab:sim_params}.

The governing equations are discretised using the finite-volume method. Temporal derivatives are integrated with a first-order implicit Euler scheme. The convective term in the momentum equation is discretised using the LUST scheme, a blend of linear and linear-upwind interpolation that combines second-order accuracy with the stability of upwind differencing, while the remaining divergence and gradient terms use second-order Gauss linear interpolation. Laplacian terms are evaluated with Gauss linear interpolation and an uncorrected surface-normal gradient, which is appropriate for the orthogonal co-located grid employed here.

On the HGD side, solid walls are treated as impenetrable barriers: if a stochastic swap would move a solid element into a wall cell or into a cell where the solid fraction has reached the critical packing value $\phi_c$, the swap is rejected and the velocity of that element is set to zero. Case-specific boundary configurations are described in the corresponding sections of the results.

\begin{table*}[t]
\centering
\small
\setlength{\tabcolsep}{4pt}
\caption{Summary of simulation parameters used in the validation cases: Case 1, single-particle settling; Case 2, hindered settling; Case 3 and Case 4, silo deposition.}
\label{tab:sim_params}
\begin{tabular}{llcccc}
\toprule
Parameter & Unit & Case 1 & Case 2 & Case 3 & Case 4 \\
& & Single-particle settling & Hindered settling & \makecell{Silo deposition\\(glass beads)} & \makecell{Silo deposition\\(alumina)} \\
\midrule
\multicolumn{6}{l}{\textit{Numerical and geometric settings}} \\
Grid cells, $N_x \times N_y$      & --            & $15 \times 13$      & $13 \times 90$      & $140 \times 156$    & $80 \times 210$ \\
Cell size, $\Delta x = \Delta y$  & m             & 0.05                & 0.004               & 0.002               & 0.0005 \\
Domain size, $L_x \times L_y$     & m             & $1.1 \times 0.65$   & $0.052 \times 0.36$ & $0.28 \times 0.31$  & $0.04 \times 0.105$ \\
Heterarchical layers, $M$         & --            & 100                 & 400                 & 200                 & 200 \\
Time-step size, $\Delta t$        & s             & $1\times10^{-4}$    & $1\times10^{-4}$    & $1\times10^{-3}$    & $1\times10^{-4}$ \\
\addlinespace
\multicolumn{6}{l}{\textit{Material and fluid properties}} \\
Particle size, $s$                & mm            & 0.5, 1.5, 2.0       & 0.001--1.0          & 0.36, 1.0           & 0.064, 0.15 \\
Particle size distribution        & --            & monodisperse        & polydisperse        & bidisperse          & bidisperse \\
Particle density, $\rho_p$        & kg\,m$^{-3}$  & 2560, 2560, 2480    & 2650                & 2550                & 3965 \\
Fluid density, $\rho_f$           & kg\,m$^{-3}$  & 997                 & 997                 & 997                 & 786 \\
Fluid viscosity, $\mu_f$          & Pa\,s         & $1.0\times10^{-3}$  & $1.0\times10^{-3}$  & $1.0\times10^{-3}$  & $2.2\times10^{-3}$ \\
\addlinespace
\multicolumn{6}{l}{\textit{HGD diffusion parameters}} \\
Critical solid fraction, $\phi_c$  & --  & 0.5  & 0.5  & 0.55 & 0.5 \\
Mixing coefficient, $\alpha_{\max}$ & --  & 0.3  & 0.3  & 0.3  & 0.3 \\
\bottomrule
\end{tabular}
\end{table*}

\section{Results}
The HGFD framework is evaluated through three test cases, designed to verify the key physical mechanisms introduced in the model. The first case (Section~\ref{sec:single_particle}) validates the inertial particle dynamics at the single-particle level against the benchmark settling experiments by Mordant and Pinton~\cite{mordant_velocity_2000}. The second case (Section~\ref{sec:hindered}) explores the model's capacity to predict collective particle motion in terms of the vertical size segregation that develops during the hindered settling experiments by Li and van Zyl~\cite{li_hindered_2022}. The third case (Section~\ref{sec:seg_types}) presents even more complex deposition scenarios involving bidisperse particles settling through a viscous fluid, providing a stringent test of the coupled framework's ability to capture fluid-driven segregation patterns observed experimentally by Athani et al.~\cite{athani_scale_2025}.

\subsection{Single particle settling: Validation against experiment}
\label{sec:single_particle}

This test provides a validation case for the HGFD formulation in a fully dynamic, inertial regime by simulating the settling of a single spherical particle in a closed container filled with water, following the benchmark experiments by Mordant and Pinton~\cite{mordant_velocity_2000}. In their study, the settling velocity of individual spherical particles with varying diameters was measured using an acoustic Doppler technique, providing high-resolution temporal records of the particle velocity throughout the settling process. This configuration involves finite Reynolds number effects, thereby assessing the model’s ability to capture transient particle acceleration and terminal settling behaviour.

\begin{figure}[tbp]
    \centering
    \includegraphics[width=\columnwidth]{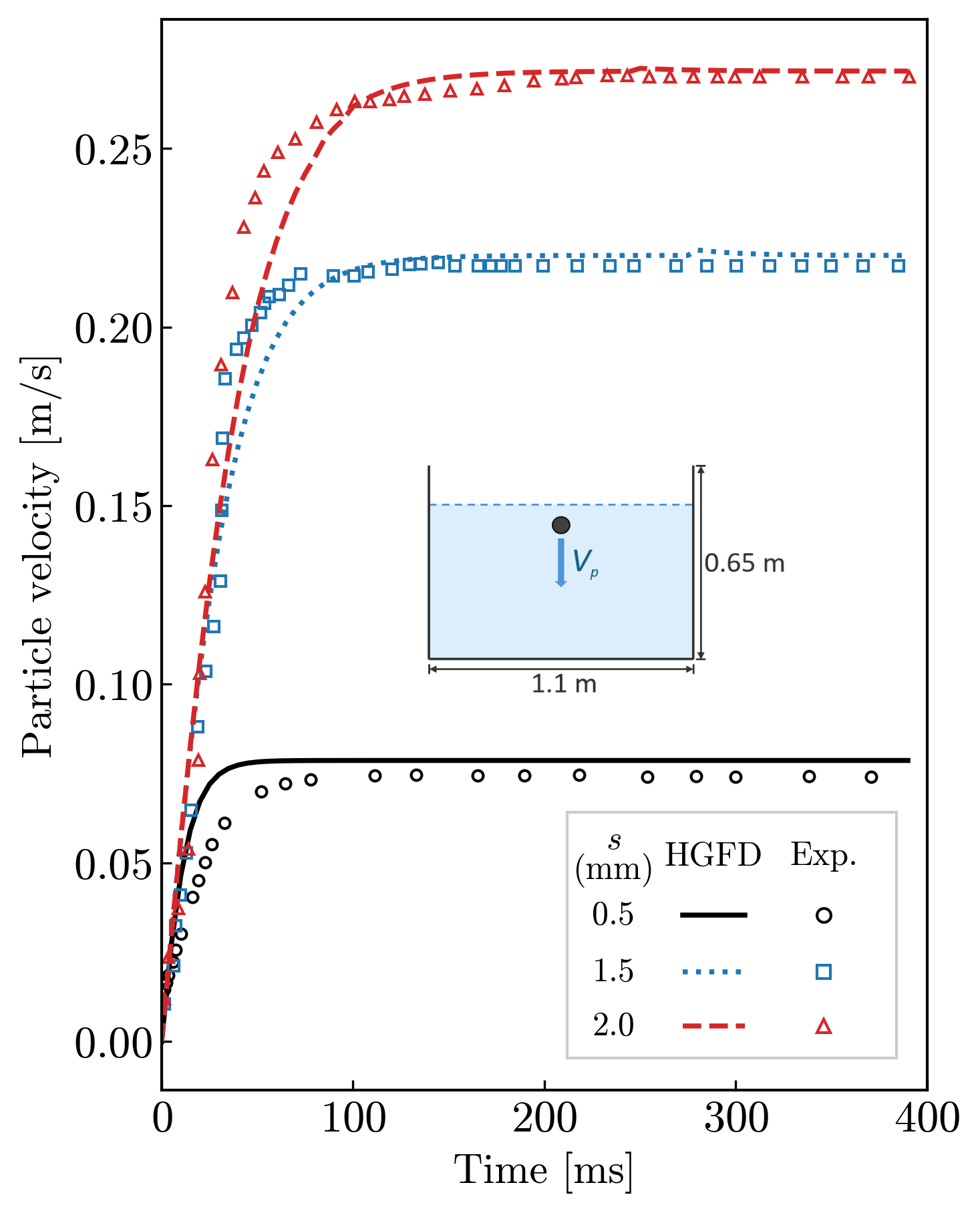}
    \caption{Single-particle settling validation against the experimental data of Mordant and Pinton~\cite{mordant_velocity_2000}. Particle settling velocity as a function of time for three particle diameters ($d_p = 0.5$, $1.5$, and $2.0$~mm).}
    \label{fig:mordant}
\end{figure}

The simulation domain and material properties are listed in Table~\ref{tab:sim_params} (Case~1). The domain geometry is shown in the inset of Fig.~\ref{fig:mordant}. The top boundary is open, and no-slip conditions are applied at the remaining walls. The particle is released from rest in a quiescent fluid, and the simulation is run until the terminal velocity is reached.

Figure~\ref{fig:mordant} compares the temporal evolution of the particle settling velocity predicted by the HGFD formulation against the experimental data for three particle diameters ($s = 0.5$, $1.5$, and $2.0$~mm). The numerical solution reproduces the acceleration from rest and the gradual approach to the terminal velocity of each particle size, showing relatively small deviations prior to the establishment of the terminal velocity. The predicted terminal velocities are in close agreement with the experimental measurements for all three particle sizes.  The results also correctly reproduce the increase in terminal velocity with particle size, consistent with the expected scaling of drag and gravitational forces.

\subsection{Hindered settling and vertical size segregation}
\label{sec:hindered}

\begin{figure}[htbp]
    \centering
    \includegraphics[width=\columnwidth]
    {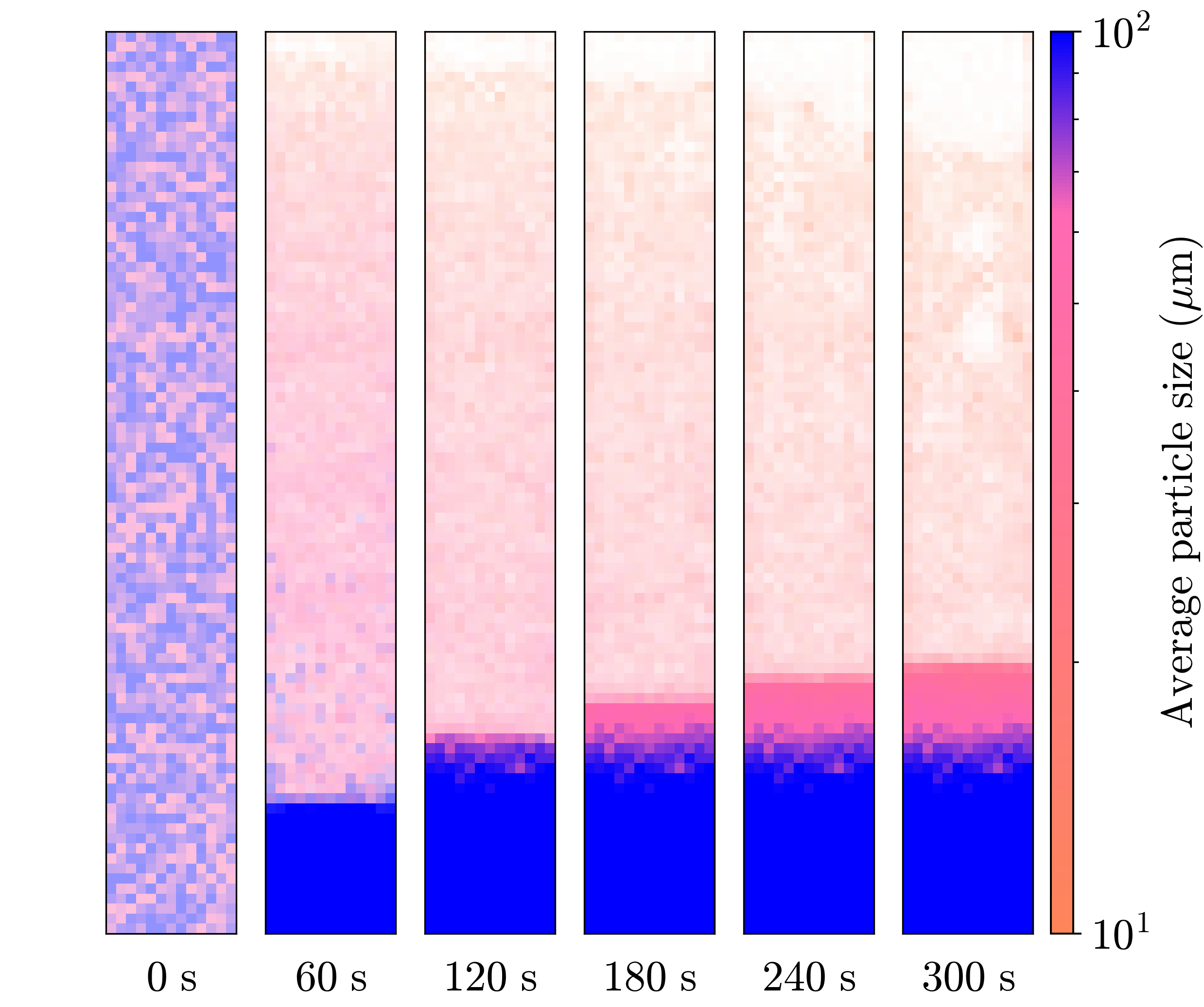}
    \caption{Temporal evolution of the predicted particle-size field during hindered settling from 0 to 300~s. Coarse particles are shown in blue and finer particles in orange.}
    \label{fig:hindered_segregation_evolution}
\end{figure}

This test evaluates the HGFD framework under hindered settling conditions, where drag depends on local solid concentration and the polydisperse suspension develops vertical size stratification over time. The case setup follows Li and van Zyl~\cite{li_hindered_2022}, who measured particle size distributions at several vertical zones of a settling column at times $t = 300$, 900, and 1800~s after stirring was ceased. The simulation domain and material properties are listed in Table~\ref{tab:sim_params} (Case~2). The simulation domain matches the experimental column dimensions (5.2~cm wide, 36~cm tall). The top boundary is open, and no-slip conditions are applied at the left, right, and bottom walls. The suspension is initialised as spatially uniform throughout the domain. 

\begin{figure}[t]
    \centering
    \includegraphics[width=\columnwidth]
    {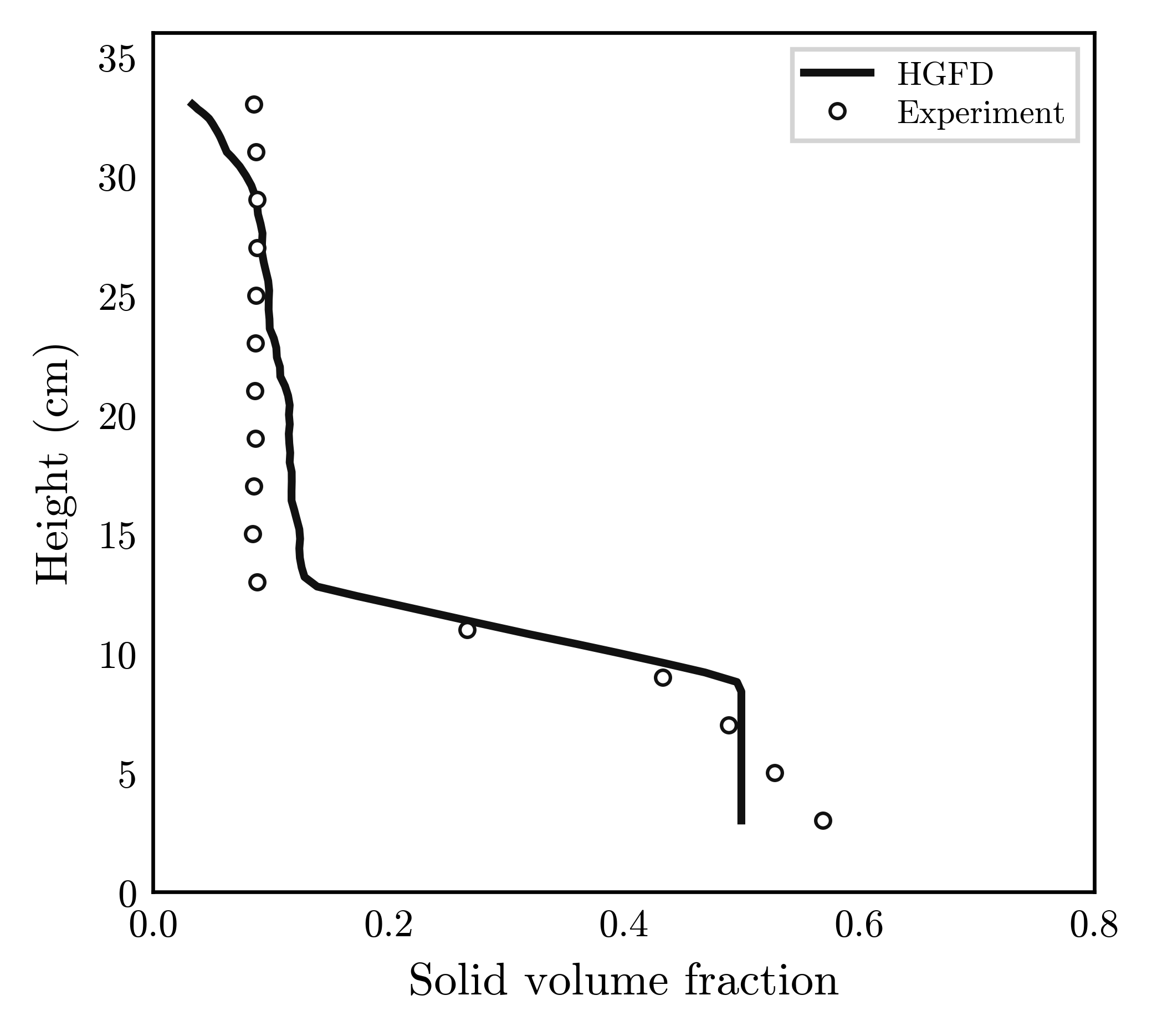}
    \caption{Comparison of predicted (HGFD) and experimental~\cite{li_hindered_2022} vertical profiles of solid volume fraction at $t = 300$~s.}
    \label{fig:hindered_solid_fraction}
\end{figure}

To reduce computational cost, the drag force is evaluated using a lower size limit: particles smaller than 25~$\mu$m are assigned an effective diameter of 25~$\mu$m for the drag force and the corresponding momentum source term returned to the fluid phase. Their stored size attributes and transport behaviour within the heterarchical lattice remain unchanged. At this size range, the drag relaxation time is orders of magnitude shorter than the advection time scale, so that these particles contribute negligibly to the interphase momentum source term compared with the coarser fraction, while imposing a prohibitively small time step for stable explicit coupling. The lower size limit removes this restriction at negligible cost to accuracy.

Figure~\ref{fig:hindered_segregation_evolution} shows the temporal evolution of the predicted particle-size field over 300~s of settling. At $t = 0$~s the suspension is spatially homogeneous. As settling proceeds, vertical stratification develops progressively: a coarse sediment layer accumulates at the base, while the mean particle size in the upper portion of the column decreases with time. By $t = 300$~s three distinct regions are visible: a fine-particle zone at the top of the column with average sizes of order 10~$\mu$m, a coarse sediment layer at the base with sizes exceeding 100~$\mu$m, and a transitional region between them in which the mean particle size varies gradually with height.

\begin{figure*}[t]
    \centering
    \includegraphics[width=0.7\linewidth]
    {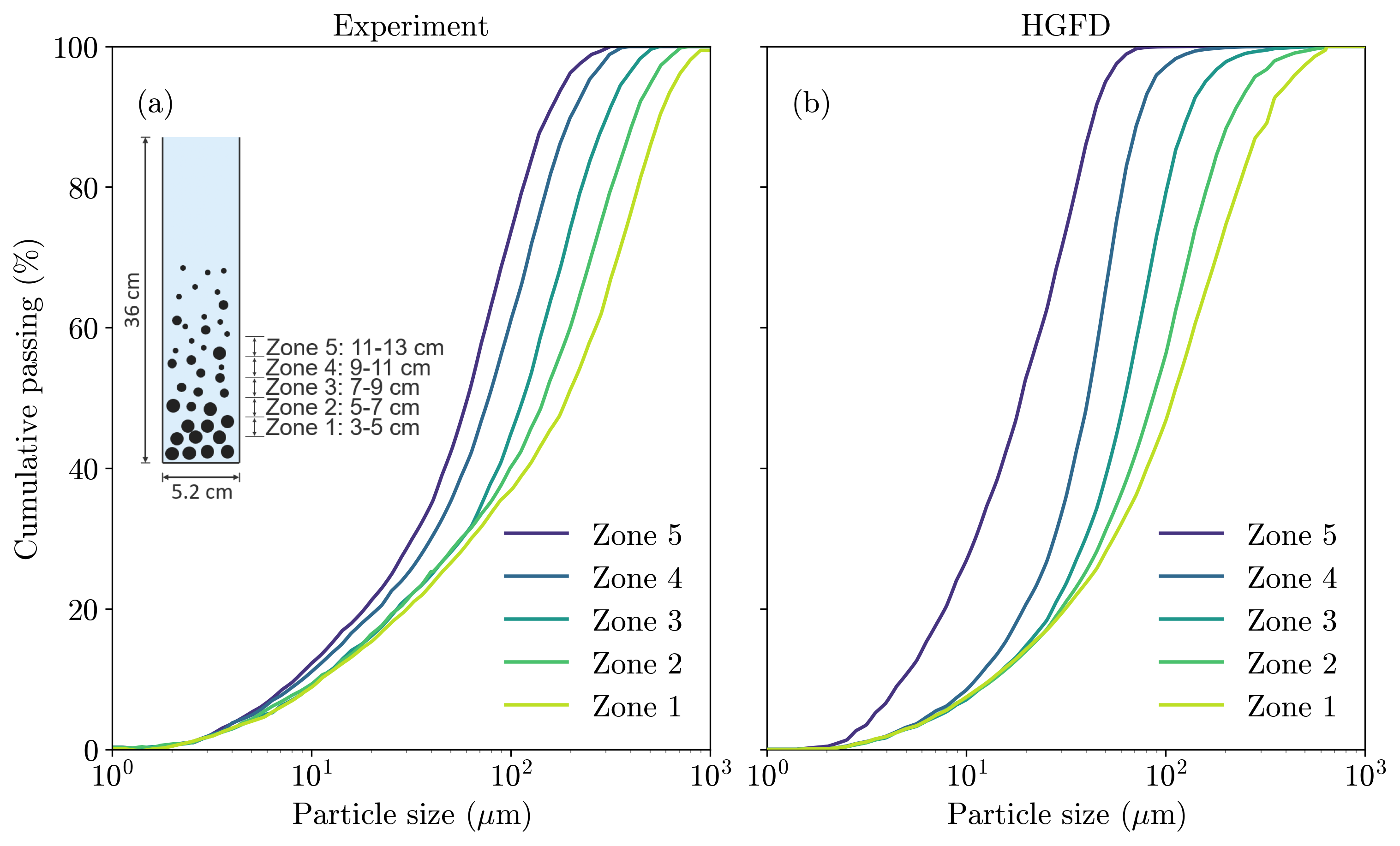}
    \caption{Comparison of predicted (HGFD) and experimental~\cite{li_hindered_2022} cumulative particle-size distributions at $t = 300$~s of settling for five vertical zones (Zone~1: 3--5~cm to Zone~5: 11--13~cm from the base).}
    \label{fig:hindered_segregation_cdf}
\end{figure*}

The comparison is made at $t = 300$~s, the earliest reported time point at which distinct size fractionation across the column height is already evident. Extension to later times is not pursued owing to the substantial computational cost over longer durations, as most of the stratification has already been  completed by this stage. Figure~\ref{fig:hindered_solid_fraction} compares the predicted and experimental vertical profiles of solid volume fraction. The predicted profile captures the overall structure observed experimentally: a dense sediment layer at the base, a sharp transition near 13~cm, and a dilute upper region with approximately uniform concentration around 0.07. Above 13~cm the predicted concentrations agree closely with the experimental data. Below this height, the model reproduces the increasing trend toward the base but with a sharper transition, as the constant critical solid fraction of 0.5 prescribed in the HGFD framework caps the local concentration. In practice, pore filling by fine particles between coarser grains can raise the packing fraction beyond this limit.

To assess this prediction quantitatively, Figure~\ref{fig:hindered_segregation_cdf} compares the predicted and experimental cumulative particle-size distributions at $t = 300$~s. Five vertical zones are selected from the experimental dataset of~\cite{li_hindered_2022} (Zone~1: 3--5~cm to Zone~5: 11--13~cm from the base), corresponding to the lower portion of the column where measurable segregation is reported. Above this range, the experimental distributions converge and show negligible height dependence. Both the experimental and predicted distributions shift progressively toward finer sizes with increasing height, and the predicted ordering of the five zones is consistent with the experimental data. The predicted distributions for Zones 1–4 show similar spacing to the experimental data, whereas Zone 5 is more clearly separated from the remaining zones than in the experiment, indicating that the model over-predicts size fractionation near the upper suspension front. This over-separation may partly reflect the idealised initial condition assumed in the simulation, where the suspension is prescribed as perfectly uniform, as well as the heightened sensitivity of the hindered settling correction at low solid fractions near the top of the suspension. In addition, the experimental sampling procedure, which involves extracting the suspension in 2~cm increments, necessarily introduces some remixing within each sampled interval, smoothing the measured distributions relative to the undisturbed suspension.
\FloatBarrier
\subsection{Segregation-type validation under fluid-dominated conditions}
\label{sec:seg_types}

The preceding tests validated vertical size stratification against experimental data. This section extends the validation to more complex deposition scenarios involving confined geometries, where both vertical and lateral segregation develop and are experimentally characterised, providing a more comprehensive test of the framework.

The experimental reference is the study of Athani et al.~\cite{athani_scale_2025}, who discharged bidisperse particle mixtures into fluid-filled silos of varying geometry and fluid properties. They observed that the resulting deposit structure depends strongly on two dimensionless groups: the silo aspect ratio $\mathrm{Ar} = H/W$, where $H$ and $W$ are the silo height and width, and the Stokes number $\mathrm{St}$ 
\begin{equation}
\mathrm{St} = \frac{\rho_p\,d^2\,v_t}{18\,\eta\,H},
\label{eq:stokes}
\end{equation}
where $\rho_p$ is the particle density, $d$ is the volume-averaged particle diameter, $v_t = (\rho_p - \rho_f)g\,d^2 / (18\,\eta)$ is the terminal settling velocity, and $\eta$ is the fluid dynamic viscosity~\cite{athani_scale_2025}. As $\mathrm{St}$ decreases, the segregation pattern transitions from laterally dominated to vertically dominated. Under weakly fluid-coupled conditions ($\mathrm{St} \gg 1$), coarse particles accumulate along the silo walls during feeding, where the heap surface forms a thin flowing layer in which shear-induced diffusion transports the coarser fraction outward toward the walls, while finer particles remain near the central feed region; as shown in Section~\ref{sec:Motivation} (Fig.~\ref{fig:motivation_patterns}), this pattern is already reproduced by the basic HGD formulation without fluid coupling. The present validation focuses on two fluid-dominated configurations ($\mathrm{St} \ll 1$) from the same experimental dataset, which require the momentum exchange introduced by the HGFD extension. The two cases differ in Stokes number, silo geometry, particle material, and surrounding fluid, thereby testing the generality of the framework across different physical settings. Table~\ref{tab:seg_cases} summarises the key parameters; complete specifications follow those reported in~\cite{athani_scale_2025}. In both configurations, the silo geometry follows the experimental setup, with no-slip conditions applied at all solid walls on the CFD side and solid boundaries prescribed at the same locations on the HGD grid.


\begin{table}[b]
  \centering
    \caption{Configurations selected for the silo deposition validation. Case~A is included for reference; its pattern is reproduced by the basic HGD model without fluid coupling (Fig.~\ref{fig:motivation_patterns}). Complete material and geometric specifications are listed in Table~\ref{tab:sim_params}.}
  \label{tab:seg_cases}
  \footnotesize
  \begin{tabular}{l l l c c c}
    \hline
    Case & Particles & Fluid & $H$\,(cm) & Ar & St \\
    \hline
    A & alumina & air         & 12  & 1.5  & 5.2 \\
    B & glass beads & soapy water & 19 & 1.3 & 0.16 \\
    C & alumina & propanol    & 6  & 1.5  & $1.7\!\times\!10^{-4}$ \\
    \hline
  \end{tabular}
\end{table}

\begin{figure*}[t]
  \centering
  \includegraphics[width=\textwidth]{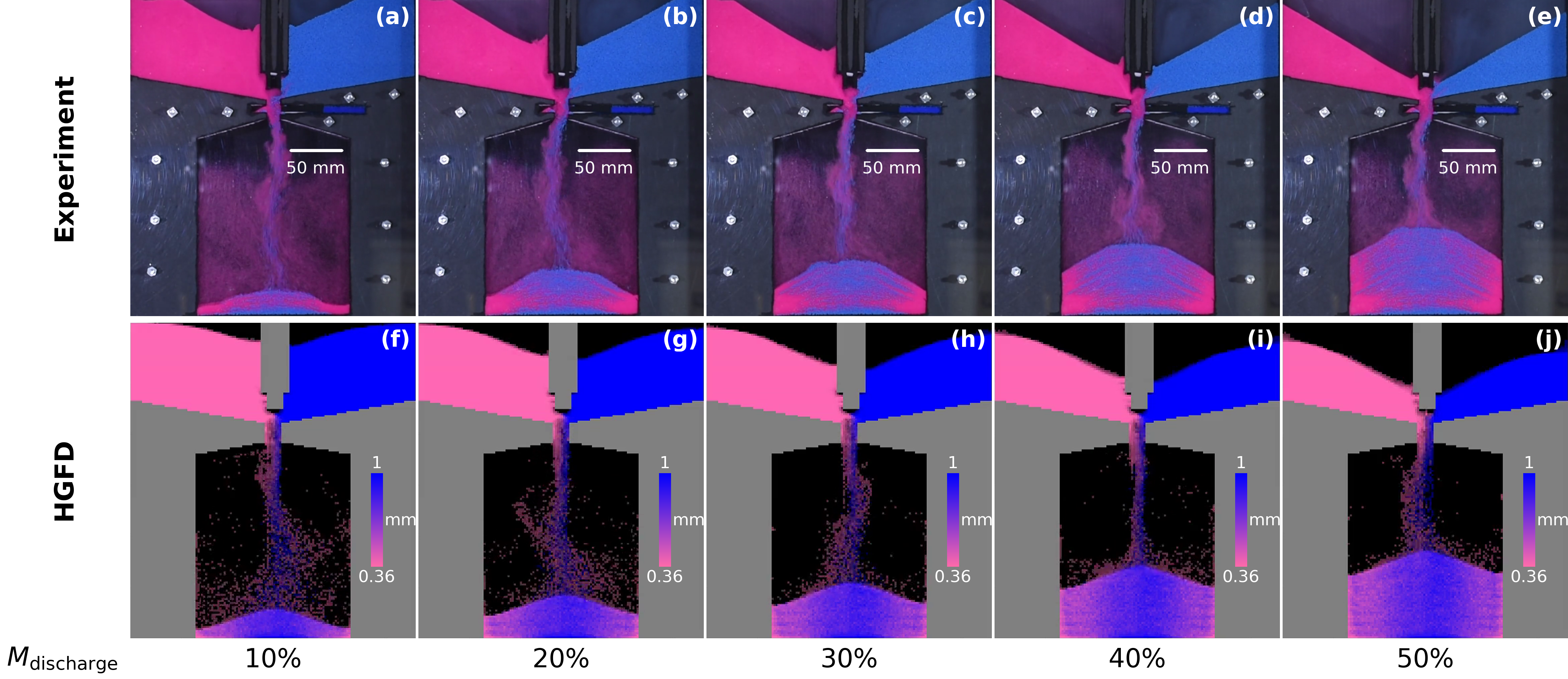}
\caption{Case~B deposition sequence (glass beads in soapy water, $\mathrm{St} = 0.16$): comparison between experimental observations adapting methodology from~\cite{athani_scale_2025}, here controlling fluxes of the two particle sizes (top row) and the HGFD predictions (bottom row) at five discharge stages.}
  \label{fig:case2_pattern}
\end{figure*}

\subsubsection{Case~B: moderate Stokes number ($\mathrm{St} = 0.16$)}
Figure~\ref{fig:case2_pattern} compares the simulated and experimental deposition sequences for Case~B (Table~\ref{tab:seg_cases}). Particles are discharged from two side-by-side hoppers---one containing the coarse fraction (blue, 1~mm) and the other the fine fraction (pink, 0.36~mm)---through a central opening into the fluid-filled silo below. To match the experimental feeding condition, the mixing coefficient $\alpha$ in the upper hoppers is calibrated so that both fractions discharge at the same rate. The figure shows five snapshots at 10\%, 20\%, 30\%, 40\%, and 50\% of the total discharged mass. The comparison is made at matched discharge fractions rather than at matched physical times, for reasons discussed at the end of this section.

In both the experiment and the HGFD simulation, the descending particle stream entrains the surrounding fluid and generates lateral recirculation on either side of the central stream. The coarse particles (blue) settle rapidly and accumulate at the base of the deposit, forming a central core, while the finer particles (pink) are carried laterally by the recirculating fluid and accumulate towards the silo walls and the upper surface of the growing heap. Throughout the discharge sequence, the deposit develops a clear centre-to-wall size gradient, and the HGFD prediction reproduces the spatial arrangement and evolution observed in the experiment at each discharge stage. At this $\mathrm{St}$, the drag relaxation time is short relative to the filling time scale, so that particle velocities remain closely tied to the local fluid motion. This allows fluid-driven lateral transport to dominate, producing a deposition structure that is qualitatively distinct from the weakly fluid-coupled Case A (Fig.~\ref{fig:motivation_patterns}).

\begin{figure}[t]
  \centering
  \includegraphics[width=\linewidth]{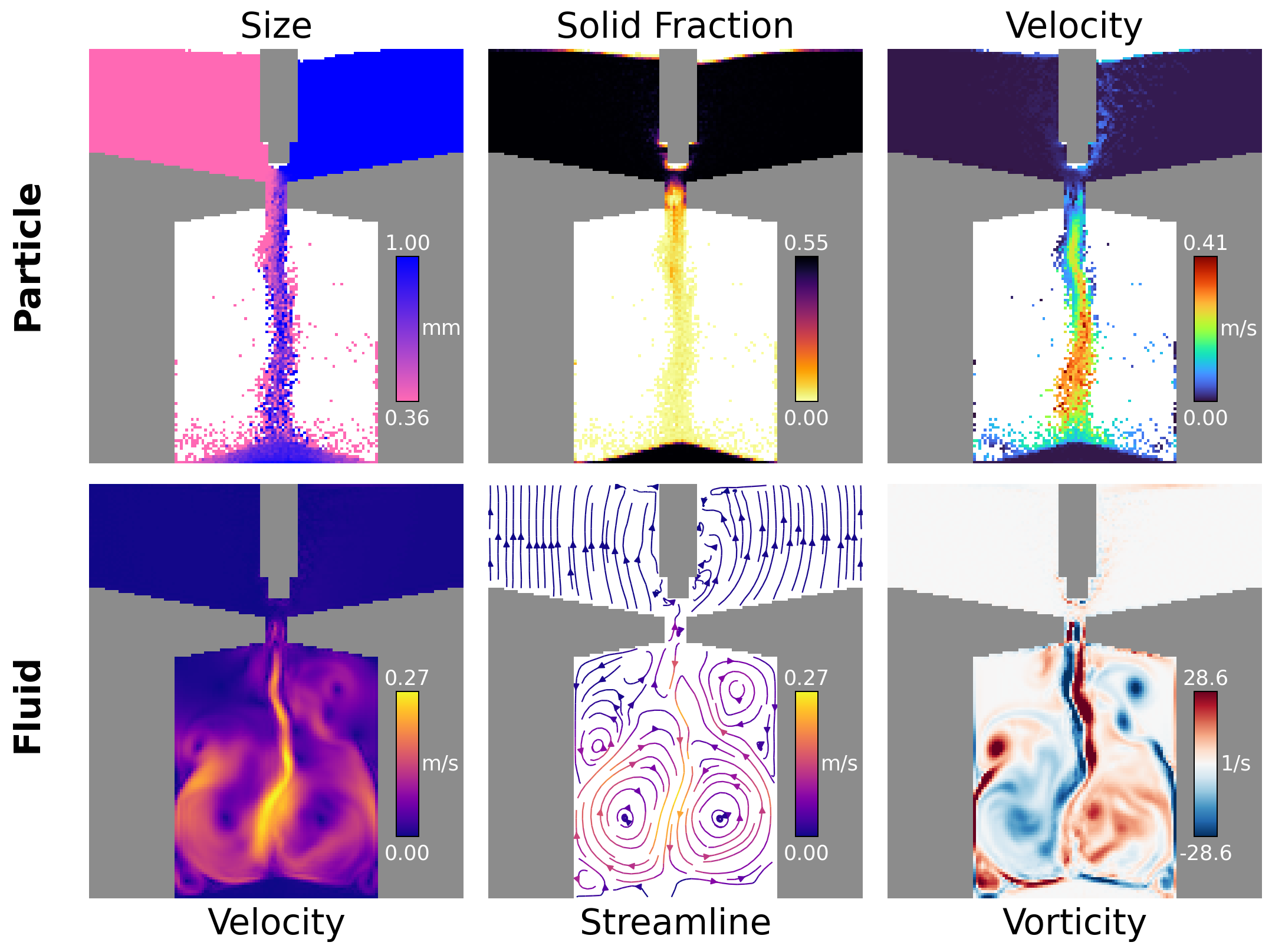}
  \caption{Detailed HGFD and fluid fields for Case~B (glass beads in soapy water, $\mathrm{St} = 0.16$), including particle size, solid fraction, particle velocity, fluid velocity, streamline pattern, and vorticity.}
  \label{fig:case2_detail}
\end{figure}

\begin{figure*}[t]
  \centering
  \includegraphics[width=\textwidth]{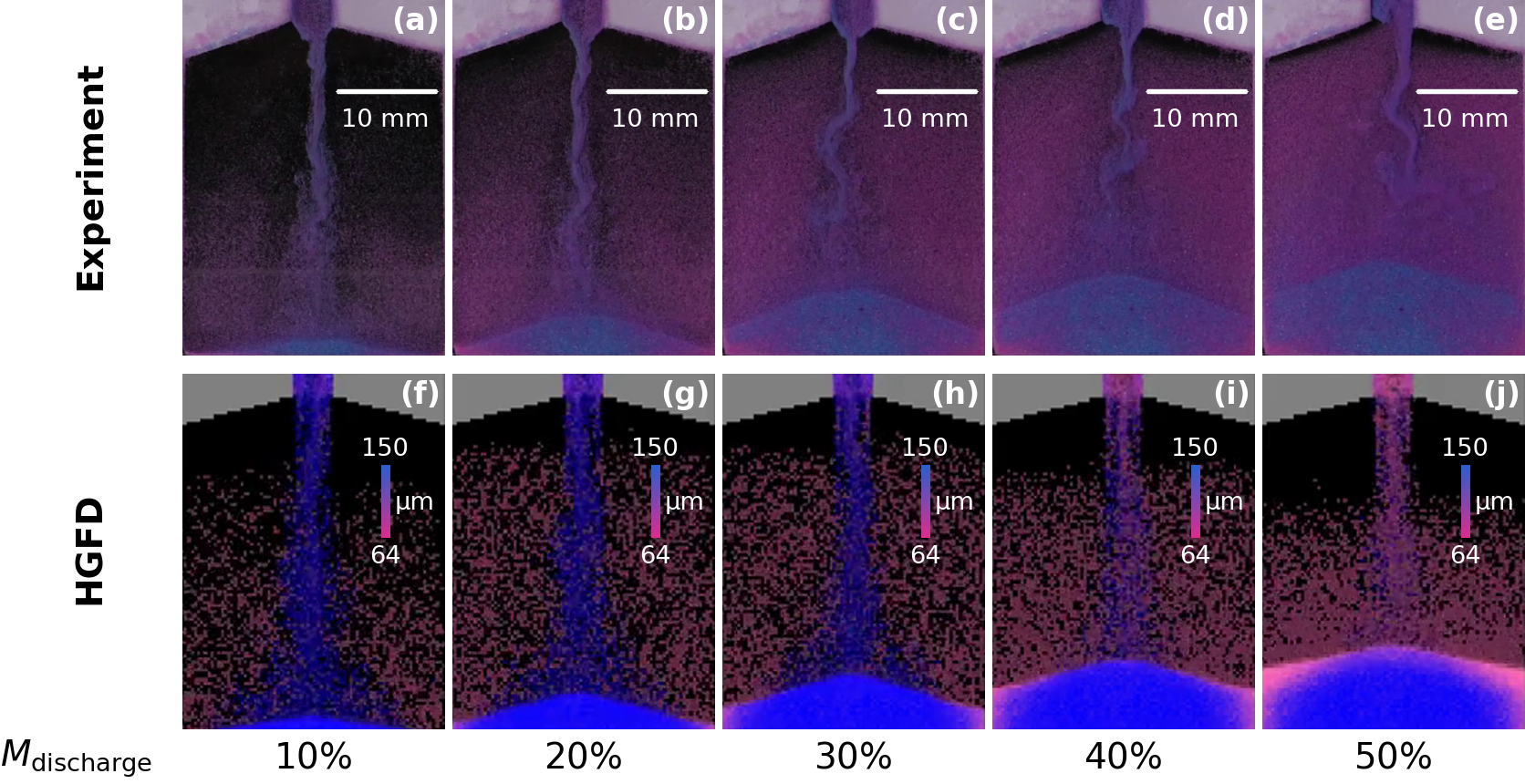}
\caption{Case~C deposition sequence (alumina in propanol, $\mathrm{St} = 1.7\times10^{-4}$): comparison between the experimental observations from~\cite{athani_scale_2025} (top row) and the HGFD predictions (bottom row) at five discharge stages.}
  \label{fig:case3_pattern}
\end{figure*}

Figure~\ref{fig:case2_detail} shows the detailed particle and fluid fields at an intermediate stage of the discharge. The solid fraction field shows high concentrations in the hoppers and the deposit at the base, while the descending particle stream in between remains relatively dilute. The particle velocity field shows the highest speeds along the central stream, decaying rapidly away from it. On the fluid side, the velocity field shows two symmetric circulation cells on either side of the central particle stream, generated as the descending particles drag the surrounding fluid downward and force a return flow along the silo walls. The streamline pattern confirms this broadly symmetric recirculation structure, and the vorticity field shows broadly antisymmetric vortex pairs consistent with the shear generated at the interface between the downward particle stream and the surrounding return flow. These recirculation cells are the mechanism responsible for the lateral size sorting observed in the deposit: fine particles are preferentially carried outward by the return flow while coarse particles settle through the central stream.

Despite the overall agreement in the segregation pattern, the simulated discharge rate from the hoppers is slower than in the experiment, because particles in the densely packed hoppers must move laterally to reach the outlet, and this lateral transport relies on the diffusion function rather than on resolved contact mechanics. This also explains why the simulated heap exhibits a slightly steeper angle of repose than the experimental deposit. These differences are consistent with the mesoscopic nature of the HGFD framework, which represents lateral transport through a diffusion function rather than resolved contact forces.

\subsubsection{Case~C: low Stokes number ($\mathrm{St} = 1.7\times10^{-4}$)}
Figure~\ref{fig:case3_pattern} presents the simulated and experimental deposition sequences for Case~C (Table~\ref{tab:seg_cases}). Unlike Case~B, the particles here are initially well-mixed in a single hopper and discharged through a central opening into the lower half of an hourglass-shaped vessel filled with propanol. The figure shows five snapshots at 10\%, 20\%, 30\%, 40\%, and 50\% of the total discharged mass.

At this much lower $\mathrm{St}$, viscous drag dominates particle inertia far more strongly than in Case~B. As in Case~B, the settling particle stream entrains the surrounding fluid and generates recirculation. However, the stronger viscous coupling means that finer particles (pink, 64~$\mu$m) are more readily carried upward by the return flow along the silo walls, while the coarse particles (blue, 150~$\mu$m) settle rapidly through the central stream to the base. The result is a vertically stratified deposit rather than the lateral centre-to-wall gradient observed in Case~B.

The HGFD prediction reproduces this pattern throughout the discharge sequence: a narrow columnar stream of coarse particles descends through the centre, and the deposit at the base develops clear vertical stratification with coarse material at the bottom and fine material above. The contrast with Case~B confirms that the framework captures the sensitivity of the segregation pattern to the Stokes number: as $\mathrm{St}$ decreases, the deposit transitions from laterally dominated to vertically dominated segregation.
\FloatBarrier

\section{Conclusion}

In this work, we extended the quasi-static heterarchical granular dynamics (HGD) framework to fluid-coupled, inertia-influenced granular systems by introducing explicit particle dynamics governed by local force balance and coupling the granular phase to a fluid-fraction-weighted incompressible fluid solver.

The use of co-located grids and a minimal set of exchanged Eulerian fields allows a transparent and computationally efficient integration of the two solvers. Probabilistic consistency of the HGD update is maintained by ensuring that the total transition probability remains bounded, and the staggered explicit coupling scheme requires only a single exchange per time step without sub-iterations.

The framework was evaluated through three progressively complex test cases. The single-particle settling test confirmed that the stochastic lattice formulation supports force-balance-driven inertial dynamics with quantitative agreement against the experimental measurements by Mordant and Pinton~\cite{mordant_velocity_2000}. The hindered-settling case showed that concentration-dependent drag and vertical size stratification develop naturally from the extended framework within physically realistic time scales. The silo deposition cases demonstrated that the framework captures the transition from laterally dominated to vertically dominated segregation as the Stokes number decreases, reproducing the experimental patterns of Athani et al.~\cite{athani_scale_2025} across different particle materials, fluids, and silo geometries. Under weakly fluid-coupled conditions ($\mathrm{St} \gg 1$), the basic HGD formulation without fluid coupling already captures the observed segregation pattern, confirming that the HGFD extension is only required when fluid effects are physically significant.

These cases address the objectives stated in Section~1: the single-particle test validates the inertial extension~(i), the stable execution of all coupled simulations confirms the robustness of the coupling strategy~(ii), and the hindered-settling and silo deposition results demonstrate the framework's capability to reproduce collective phenomena across multiple regimes~(iii). Taken together, the results show that the proposed HGFD framework retains the mesoscopic transport structure of HGD while extending its applicability to fluid-coupled granular systems.

The present study is limited to laminar flow conditions. The mesoscopic treatment of inter-particle interactions through a diffusion function, rather than resolved contact mechanics, leads to discrepancies in hopper discharge rates and angle of repose, as discussed in Section~\ref{sec:seg_types}. In addition, all simulations presented here are two-dimensional; extension to three-dimensional geometries is straightforward in principle but has not yet been tested. The computational cost of the framework has not been formally benchmarked against established methods such as CFD--DEM, although the cell-based stochastic formulation is expected to offer significant advantages for large-scale systems. Future work will investigate extensions to higher Reynolds number regimes, the incorporation of more detailed discharge models, and systematic calibration against industrial-scale datasets. The framework provides a foundation for scalable simulation of industrial granular--fluid processes, including storage, conveying, and multiphase deposition systems.
\FloatBarrier
\clearpage

\begin{appendices}
\section{Continuum limit of the mass conservation equation}
\label{app:mass}
This appendix presents the discrete update rule for void transport on the heterarchical lattice and its continuum limit, extending the derivation in~\cite{marks_heterarchical_2025} to include directional advection and concentration-dependent diffusion.

Using the definitions of \(M\) and \(M_s\) introduced above, the number of
void-occupied internal coordinates at spatial position \((i,j)\) is
\begin{equation}
N_{i,j}=M-M_{s,i,j}.
\end{equation}

To examine the continuum limit of the layer-wise stochastic update, 
the discrete rules are expressed in terms of cell-level quantities 
by averaging over all internal coordinates $k$. In this formal 
reduction, the exchange factor at a given neighbour is approximated 
by the cell-level solid fraction $(1 - N/M)$, corresponding to a 
mean-field treatment in which correlations between occupancy states 
at different positions and across layers are neglected. The resulting 
cell-level update rule and its continuum limit are presented below; 
the actual computation is performed stochastically on the full 
layer-resolved lattice. The evolution of $N_{i,j}$ over one time step is then written as:
\begin{align}\label{eq:conservation_dense_adv_xy}
N_{i,j}^{t+\Delta t} 
= N_{i,j}^{t} 
+ G_{\text{diff}} 
+ G_{\text{adv}} 
- L_{\text{diff}}
- L_{\text{adv}},
\end{align}
where:
\begin{itemize}
  \item \(G_{\text{diff}}\): total diffusive gain,
  \item \(L_{\text{diff}}\): total diffusive loss,
  \item \(G_{\text{adv}}\): total advective gain,
  \item \(L_{\text{adv}}\): total advective loss.
\end{itemize}

The total diffusive change in the $x$-direction includes both leftward and rightward exchanges. Cell $(i,j)$ gains voids when a void from either neighbour $(i\pm1,j)$ exchanges with a solid-occupied coordinate at $(i,j)$, giving the factor $(1 - N_{i,j}/M)$. Conversely, cell $(i,j)$ loses voids when a void at $(i,j)$ exchanges with a solid-occupied coordinate in either neighbouring cell $(i\pm1,j)$, giving the factor $(1 - N_{i\pm1,j}/M)$. It can be expressed as:
\begin{align}
G_{\text{diff},x} &= 
P_{\text{diff},i+1,j}^t N_{i+1,j}^t 
\left(1 - \frac{N_{i,j}^t}{M} \right) \notag \\
&\quad + P_{\text{diff},i-1,j}^t N_{i-1,j}^t 
\left(1 - \frac{N_{i,j}^t}{M} \right) \label{eq:Gdiffx} \\
L_{\text{diff},x} &= 
P_{\text{diff},i,j}^t N_{i,j}^t 
\left(1 - \frac{N_{i+1,j}^t}{M} \right) \notag \\
&\quad + P_{\text{diff},i,j}^t N_{i,j}^t 
\left(1 - \frac{N_{i-1,j}^t}{M} \right). \label{eq:Ldiffx}
\end{align}
In the present formulation, diffusion is applied only in the $x$-direction, consistent with the lateral mixing mechanism described in Section~\ref{sec:fluid-coupled-hgd}.

Using similar logic, the advective change is written in terms of directional exchange probabilities. Here \(P_{x,+}\), \(P_{x,-}\), \(P_{y,+}\), and
\(P_{y,-}\) denote the probabilities of advective exchange in the \(+x\),
\(-x\), \(+y\), and \(-y\) directions, respectively. The total advective
gain and loss are then:

\begin{align}
G_{\mathrm{adv}} &=
P_{x,+,i-1,j}^{t} N_{i-1,j}^{t}
\left(1-\frac{N_{i,j}^{t}}{M}\right) \notag \\
&\quad+
P_{x,-,i+1,j}^{t} N_{i+1,j}^{t}
\left(1-\frac{N_{i,j}^{t}}{M}\right) \notag \\
&\quad+
P_{y,+,i,j-1}^{t} N_{i,j-1}^{t}
\left(1-\frac{N_{i,j}^{t}}{M}\right) \notag \\
&\quad+
P_{y,-,i,j+1}^{t} N_{i,j+1}^{t}
\left(1-\frac{N_{i,j}^{t}}{M}\right).
\label{eq:Gadv}
\end{align}

\begin{align}
L_{\mathrm{adv}} &=
P_{x,+,i,j}^{t} N_{i,j}^{t}
\left(1-\frac{N_{i+1,j}^{t}}{M}\right) \notag \\
&\quad+
P_{x,-,i,j}^{t} N_{i,j}^{t}
\left(1-\frac{N_{i-1,j}^{t}}{M}\right) \notag \\
&\quad+
P_{y,+,i,j}^{t} N_{i,j}^{t}
\left(1-\frac{N_{i,j+1}^{t}}{M}\right) \notag \\
&\quad+
P_{y,-,i,j}^{t} N_{i,j}^{t}
\left(1-\frac{N_{i,j-1}^{t}}{M}\right).
\label{eq:Ladv}
\end{align}

The directional probabilities are related to the local advective velocities by
\begin{equation}
P_{x,+,i,j}^{t}=\frac{(u_{x,i,j}^{t})^{+}\Delta t}{\Delta x},
\qquad
P_{x,-,i,j}^{t}=\frac{(u_{x,i,j}^{t})^{-}\Delta t}{\Delta x},
\end{equation}
\begin{equation}
P_{y,+,i,j}^{t}=\frac{(u_{y,i,j}^{t})^{+}\Delta t}{\Delta y},
\qquad
P_{y,-,i,j}^{t}=\frac{(u_{y,i,j}^{t})^{-}\Delta t}{\Delta y},
\end{equation}
where
\begin{equation}
(u)^+=\max(u,0),
\qquad
(u)^-=\max(-u,0).
\end{equation}
The cell-level diffusion probability is $P_{\mathrm{diff},i,j} = D_{i,j}\,\Delta t / \Delta x^2$, with the diffusivity defined as the layer average
\begin{equation}
D_{i,j} = \alpha(\phi_{i,j})\,\overline{s_k\,|u_{y,k}|}_{i,j},
\end{equation}
where $\overline{(\cdot)} = \frac{1}{M}\sum_{k}(\cdot)_k$.

In the continuum limit, the discrete update rule (Eq.~\ref{eq:conservation_dense_adv_xy}) is expanded using a Taylor series about position $(i,j)$ to second-order accuracy in the spatial spacing and first-order in $\Delta t$. Collecting terms and taking the limit as $\Delta x$, $\Delta y$, and $\Delta t$ approach zero yields the macroscopic conservation equation for the void fraction \(n = N/M\). The resulting partial differential equation incorporates both nonlinear advection and second-order diffusion:
\begin{align}
\frac{\partial n}{\partial t}
&= \frac{\partial}{\partial x}\!\left[
    D\,\frac{\partial n}{\partial x}
    + n\phi\,\frac{\partial D}{\partial x}
  \right]
\notag\\
&\quad
- \frac{\partial}{\partial x}\!\left[ u_x\,n\phi \right]
- \frac{\partial}{\partial y}\!\left[ u_y\,n\phi \right].
\end{align}
We now describe the conservation of momentum for the solid phase in the heterarchical granular framework. Both the discrete update rule and the corresponding macroscopic formulation are presented below.

\section{Continuum limit of the momentum conservation equation}
\label{app:momentum}
In the heterarchical granular dynamics (HGD) framework, momentum is exclusively carried by the solid phase. The solid-phase momentum is defined as \(\phi \vec{u}\). Momentum evolves through advection between neighbouring cells, modulated by the availability of voids in the receiving cell.

The probability of transfer is governed by the directional advection probabilities $P_{x,+}$, $P_{x,-}$, $P_{y,+}$, and $P_{y,-}$ defined in Appendix~\ref{app:mass}. Diffusive exchanges, being symmetric, do not contribute to net momentum transfer and are thus omitted. External forces such as gravity and drag are included as explicit source terms. The following update rules for the $x$- and $y$-components account for gain and loss of momentum due to neighbouring advection and local forcing terms.
\begin{align}
(\phi\, u_x)_{i,j}^{t+\Delta t} 
&= (\phi\, u_x)_{i,j}^{t} \notag \\
&+ P_{x,+,i-1,j}^t\,\phi_{i-1,j}^t\,u_{x,i-1,j}^t \left(1 - \frac{N_{i,j}^t}{M}\right) \notag \\
&+ P_{x,-,i+1,j}^t\,\phi_{i+1,j}^t\,u_{x,i+1,j}^t \left(1 - \frac{N_{i,j}^t}{M}\right) \notag \\
&+ P_{y,+,i,j-1}^t\,\phi_{i,j-1}^t\,u_{x,i,j-1}^t \left(1 - \frac{N_{i,j}^t}{M}\right) \notag \\
&+ P_{y,-,i,j+1}^t\,\phi_{i,j+1}^t\,u_{x,i,j+1}^t \left(1 - \frac{N_{i,j}^t}{M}\right) \notag \\
&- P_{x,+,i,j}^t\,\phi_{i,j}^t\,u_{x,i,j}^t \left(1 - \frac{N_{i+1,j}^t}{M}\right) \notag \\
&- P_{x,-,i,j}^t\,\phi_{i,j}^t\,u_{x,i,j}^t \left(1 - \frac{N_{i-1,j}^t}{M}\right) \notag \\
&- P_{y,+,i,j}^t\,\phi_{i,j}^t\,u_{x,i,j}^t \left(1 - \frac{N_{i,j+1}^t}{M}\right) \notag \\
&- P_{y,-,i,j}^t\,\phi_{i,j}^t\,u_{x,i,j}^t \left(1 - \frac{N_{i,j-1}^t}{M}\right) \notag \\
&+ \phi_{i,j}^t\,F_{x,i,j}^t\,\Delta t,
\end{align}
\begin{align}
(\phi\, u_y)_{i,j}^{t+\Delta t} 
&= (\phi\, u_y)_{i,j}^{t} \notag \\
&+ P_{x,+,i-1,j}^t\,\phi_{i-1,j}^t\,u_{y,i-1,j}^t \left(1 - \frac{N_{i,j}^t}{M}\right) \notag \\
&+ P_{x,-,i+1,j}^t\,\phi_{i+1,j}^t\,u_{y,i+1,j}^t \left(1 - \frac{N_{i,j}^t}{M}\right) \notag \\
&+ P_{y,+,i,j-1}^t\,\phi_{i,j-1}^t\,u_{y,i,j-1}^t \left(1 - \frac{N_{i,j}^t}{M}\right) \notag \\
&+ P_{y,-,i,j+1}^t\,\phi_{i,j+1}^t\,u_{y,i,j+1}^t \left(1 - \frac{N_{i,j}^t}{M}\right) \notag \\
&- P_{x,+,i,j}^t\,\phi_{i,j}^t\,u_{y,i,j}^t \left(1 - \frac{N_{i+1,j}^t}{M}\right) \notag \\
&- P_{x,-,i,j}^t\,\phi_{i,j}^t\,u_{y,i,j}^t \left(1 - \frac{N_{i-1,j}^t}{M}\right) \notag \\
&- P_{y,+,i,j}^t\,\phi_{i,j}^t\,u_{y,i,j}^t \left(1 - \frac{N_{i,j+1}^t}{M}\right) \notag \\
&- P_{y,-,i,j}^t\,\phi_{i,j}^t\,u_{y,i,j}^t \left(1 - \frac{N_{i,j-1}^t}{M}\right) \notag \\
&+ \phi_{i,j}^t\,F_{y,i,j}^t\,\Delta t.
\end{align}
In the continuum limit, these equations yield the macroscopic momentum conservation laws for the solid phase:

\begin{align}
\frac{\partial(\phi\, u_x)}{\partial t} 
&= -\frac{\partial}{\partial x}(\phi\, u_x^2\, n)
   -\frac{\partial}{\partial y}(u_y\, \phi\, u_x\, n) + \phi F_x, \\[8pt]
\frac{\partial(\phi\, u_y)}{\partial t} 
&= -\frac{\partial}{\partial x}(u_x\, \phi\, u_y\, n)
   -\frac{\partial}{\partial y}(\phi\, u_y^2\, n) + \phi F_y.
\end{align}

The discrete update rules presented above are the equations actually 
integrated in the stochastic HGD solver. The continuum-limit PDEs are 
not solved directly, but are derived here to confirm that the stochastic lattice formulation recovers the expected macroscopic conservation laws.
\section{Sensitivity to numerical parameters}
\label{app:convergence}

The sensitivity of the HGFD solution to three numerical parameters is examined using the hindered settling configuration (Case~2, Table~\ref{tab:sim_params}): the time step size $\Delta t$, the spatial resolution $\Delta x$, and the number of internal coordinates $M$. In each study, one parameter is varied while the other two are held at their reference values. The metric used for comparison is the cumulative particle size distribution computed over the entire simulation domain between 3 and 13~cm from the base, sampled at $t = 10$~s. This metric is chosen because the primary output of the HGFD framework is the spatial evolution of particle size distributions, which is the quantity most relevant to industrial applications such as silo segregation and granular deposition.

\subsection{Time step size}
The time step is varied over two orders of magnitude, from $\Delta t = 10^{-2}$ to $10^{-4}$~s, with the finest case $\Delta t = 5\times10^{-5}$~s used as the reference. The resulting cumulative particle size distributions are virtually indistinguishable across the entire tested range. This insensitivity is attributed to the exponential time integration (Eq.~\ref{eq:exact}), which remains stable across stiff drag regimes without requiring small time steps. The value $\Delta t = 10^{-4}$~s adopted for the hindered settling simulation is therefore well within the converged range.

\subsection{Spatial resolution}
\begin{figure}[t]
    \centering
    \includegraphics[width=\linewidth]{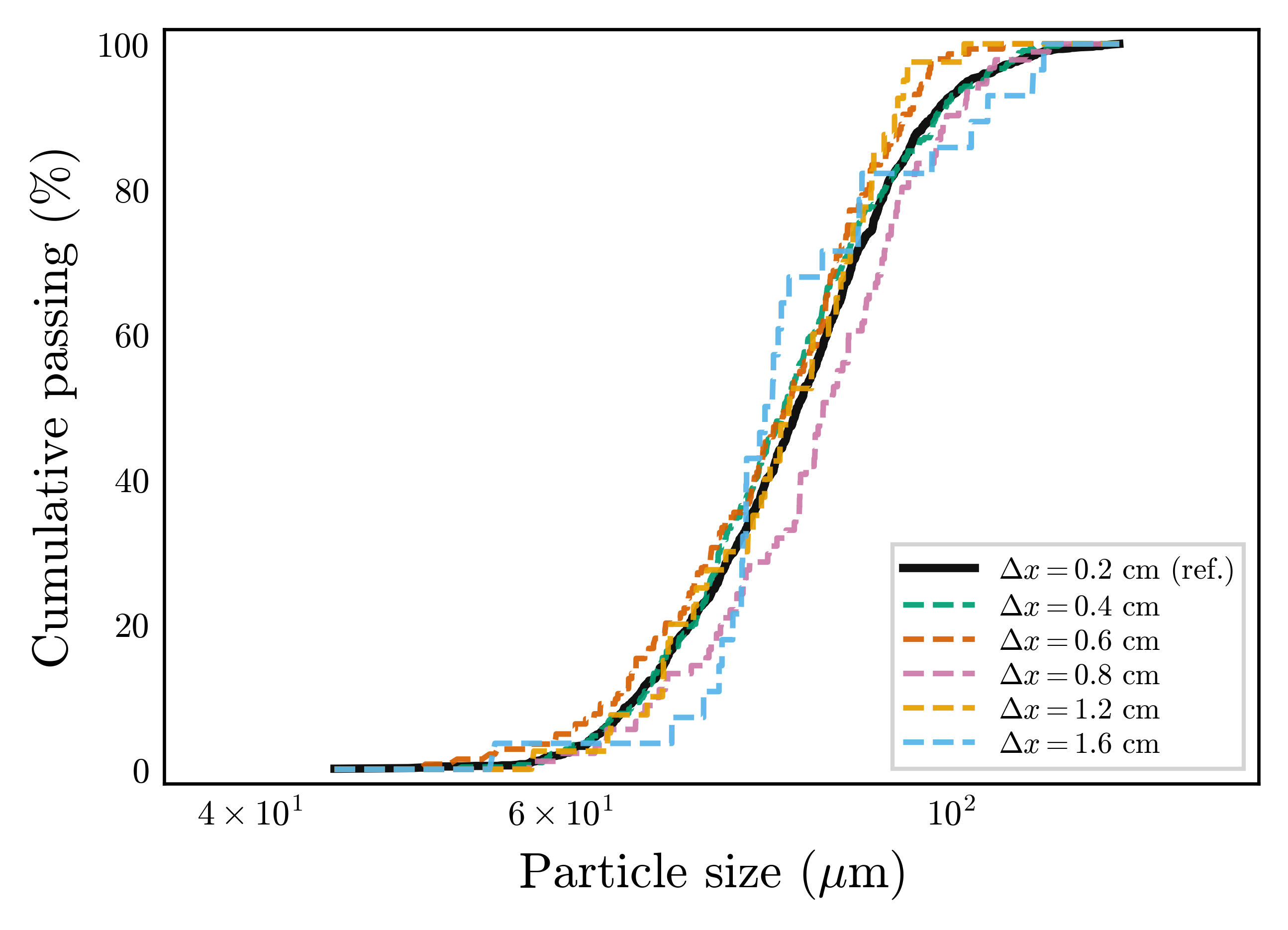}
    \caption{Sensitivity to spatial resolution: cumulative particle size distributions for varying $\Delta x$.}
    \label{fig:conv_dx}
\end{figure}

Figure~\ref{fig:conv_dx} shows the cumulative particle size distributions for six spatial resolutions ranging from $\Delta x = 0.2$ to 1.6~cm. The distributions progressively converge as $\Delta x$ decreases, with the coarsest case ($\Delta x = 1.6$~cm) showing visible deviation from the finer resolutions. For $\Delta x \leq 0.4$~cm, the distributions are virtually indistinguishable. The value $\Delta x = 0.4$~cm adopted for the hindered settling simulation (Table~\ref{tab:sim_params}) therefore provides a well-converged solution.

\subsection{Number of internal coordinates}
\begin{figure}[t]
    \centering
    \includegraphics[width=\linewidth]{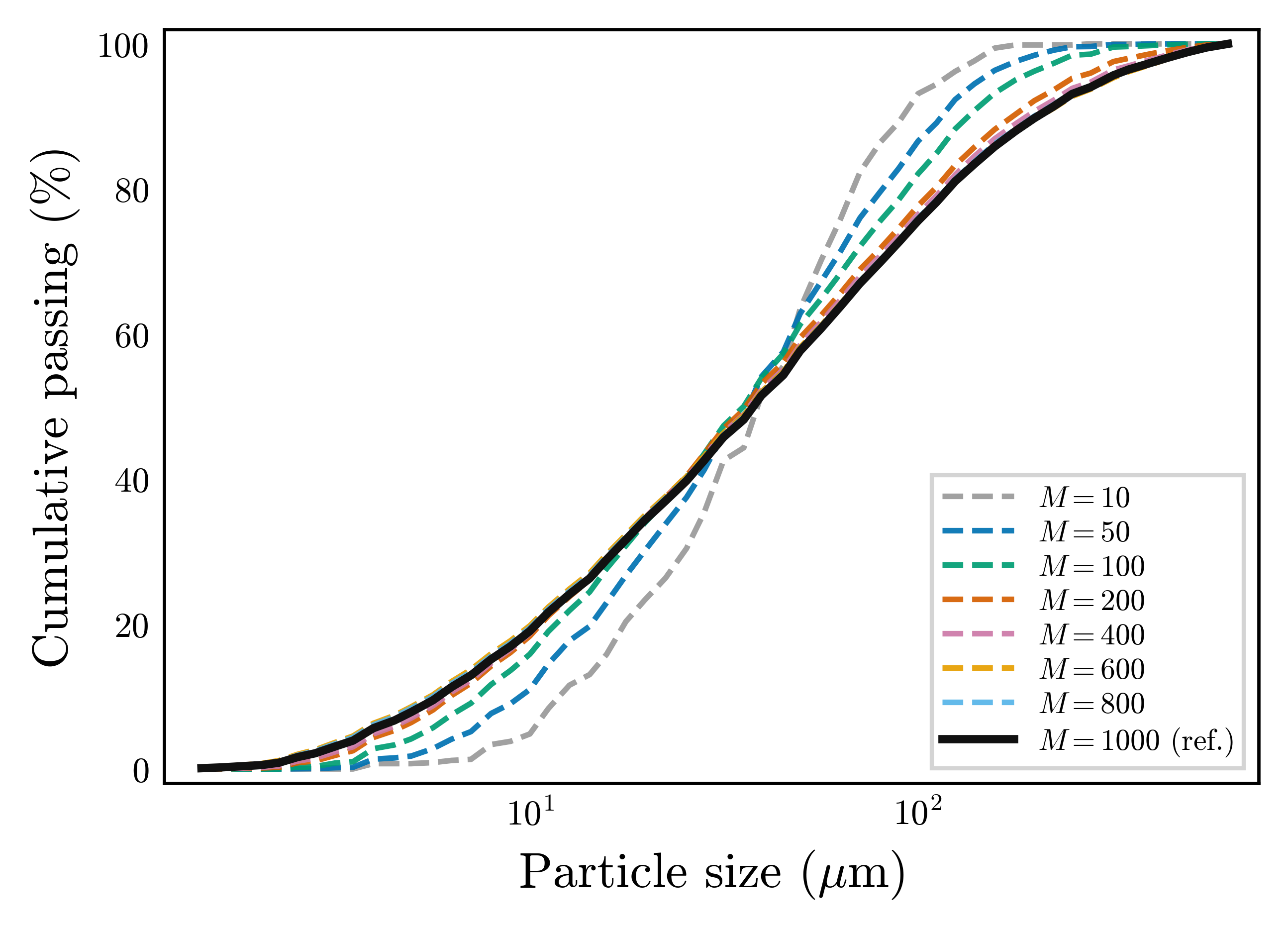}
    \caption{Sensitivity to the number of internal coordinates $M$: cumulative particle size distributions for varying $M$.}
    \label{fig:conv_M}
\end{figure}

Figure~\ref{fig:conv_M} shows the cumulative particle size distributions for $M = 20$, 50, 100, 200, 400, 600, 800, and 1000. The distributions converge rapidly with increasing $M$: the $M = 20$ case deviates visibly from the finer cases, while for $M \geq 400$ the distributions are effectively indistinguishable. The value $M = 400$ adopted for the hindered settling simulation (Table~\ref{tab:sim_params}) is well within the converged range.
In summary, the solution is insensitive to the time step size over the tested range owing to the exponential time integration scheme. Spatial resolution and the number of internal coordinates both exhibit clear convergence, with the adopted values ($\Delta x = 0.4$~cm, $M = 400$) lying well within the converged range. The hindered settling case is selected for this study as it involves the broadest particle size distribution and the strongest concentration-dependent drag effects among the three validation cases, making it the most demanding configuration for numerical resolution.

\end{appendices}

\printbibliography

@article{goz_study_2004,
	title = {Study of the numerical instabilities in {Lagrangian} tracking of bubbles and particles in two-phase flow},
	volume = {28},
	issn = {0098-1354},
	url = {https://www.sciencedirect.com/science/article/pii/S0098135404002273},
	doi = {10.1016/j.compchemeng.2004.07.035},
	number = {12},
	urldate = {2026-04-23},
	journal = {Computers \& Chemical Engineering},
	author = {G\"{o}z, M. F. and La\'{i}n, S. and Sommerfeld, M.},
	month = nov,
	year = {2004},
	pages = {2727--2733},
}

@article{guazzelli_rheology_2018,
	title = {Rheology of dense granular suspensions},
	volume = {852},
	url = {https://www.cambridge.org/core/journals/journal-of-fluid-mechanics/article/rheology-of-dense-granular-suspensions/4F792CE372121D52299422BAEADCDE74},
	urldate = {2026-02-27},
	journal = {Journal of Fluid Mechanics},
	publisher = {Cambridge University Press},
	author = {Guazzelli, \'{E}lisabeth and Pouliquen, Olivier},
	year = {2018},
	pages = {P1},
}

@article{lun_kinetic_1984,
	title = {Kinetic theories for granular flow: inelastic particles in {Couette} flow and slightly inelastic particles in a general flowfield},
	volume = {140},
	shorttitle = {Kinetic theories for granular flow},
	url = {https://www.cambridge.org/core/journals/journal-of-fluid-mechanics/article/kinetic-theories-for-granular-flow-inelastic-particles-in-couette-flow-and-slightly-inelastic-particles-in-a-general-flowfield/0BD9AC7975884A5D5F360188D647F2B6},
	urldate = {2026-02-27},
	journal = {Journal of fluid mechanics},
	publisher = {Cambridge University Press},
	author = {Lun, C. K. K and Savage, Stuart B. and Jeffrey, D. J. and Chepurniy, Nicholas},
	year = {1984},
	pages = {223--256},
}

@article{weller_tensorial_1998,
	title = {A tensorial approach to computational continuum mechanics using object-oriented techniques},
	volume = {12},
	issn = {0894-1866},
	url = {https://doi.org/10.1063/1.168744},
	doi = {10.1063/1.168744},
	number = {6},
	urldate = {2026-04-29},
	journal = {Computer in Physics},
	author = {Weller, H. G. and Tabor, G. and Jasak, H. and Fureby, C.},
	month = nov,
	year = {1998},
	pages = {620--631},
}

@article{shirolkar_fundamental_1996,
	title = {Fundamental aspects of modeling turbulent particle dispersion in dilute flows},
	volume = {22},
	issn = {0360-1285},
	url = {https://www.sciencedirect.com/science/article/pii/S0360128596000068},
	doi = {10.1016/S0360-1285(96)00006-8},
	number = {4},
	urldate = {2026-04-23},
	journal = {Progress in Energy and Combustion Science},
	author = {Shirolkar, J. S. and Coimbra, C. F. M. and Queiroz McQuay, M.},
	month = jan,
	year = {1996},
	pages = {363--399},
}

@article{iverson_physics_1997,
	title = {The physics of debris flows},
	volume = {35},
	copyright = {http://onlinelibrary.wiley.com/termsAndConditions\#vor},
	issn = {8755-1209, 1944-9208},
	url = {https://agupubs.onlinelibrary.wiley.com/doi/10.1029/97RG00426},
	doi = {10.1029/97RG00426},
	language = {en},
	number = {3},
	urldate = {2026-04-22},
	journal = {Reviews of Geophysics},
	author = {Iverson, Richard M.},
	month = aug,
	year = {1997},
	pages = {245--296},
}

@book{kunii_fluidization_2013,
	title = {Fluidization engineering},
	url = {https://books.google.com/books?hl=zh-TW&lr=&id=sGkvBQAAQBAJ&oi=fnd&pg=PP1&dq=luidization+Engineering&ots=jQPfUltBnH&sig=hfHkDBIouMq_cmuawbfGno0EGpM},
	urldate = {2026-04-22},
	publisher = {Elsevier},
	author = {Kunii, Daizo and Levenspiel, Octave},
	year = {2013},
}

@book{crowe_multiphase_2011,
	address = {Boca Raton},
	edition = {2},
	title = {Multiphase {Flows} with {Droplets} and {Particles}},
	isbn = {978-0-429-10639-2},
	doi = {10.1201/b11103},
	publisher = {CRC Press},
	author = {Crowe, Clayton T. and Schwarzkopf, John D. and Sommerfeld, Martin and Tsuji, Yutaka},
	month = aug,
	year = {2011},
}

@article{schiller_uber_1933,
	title = {Uber die grundlegenden {Berechnungen} bei der {Schwerkraftaufbereitung}},
	volume = {77},
	url = {https://cir.nii.ac.jp/crid/1570291225431729792},
	urldate = {2026-04-21},
	journal = {Z. Vereines Deutscher Inge.},
	author = {Schiller, Von L.},
	year = {1933},
	pages = {318--321},
}

@book{gidaspow_multiphase_1994,
	title = {Multiphase flow and fluidization: continuum and kinetic theory descriptions},
	shorttitle = {Multiphase flow and fluidization},
	url = {https://books.google.com/books?hl=en&lr=&id=DTAFZ9rIfQwC&oi=fnd&pg=PP2&dq=info:uNWuKn0y-YsJ:scholar.google.com&ots=cuwBv0Yblj&sig=JJaE8ZrO5BH6xYmDx2y1C6Zvi5M},
	urldate = {2026-04-21},
	publisher = {Academic press},
	author = {Gidaspow, Dimitri},
	year = {1994},
}

@article{ergun_fluid_1952,
	title = {Fluid {Flow} {Through} {Packed} {Columns}},
	volume = {48},
	url = {https://cir.nii.ac.jp/crid/1572261550410403712},
	number = {2},
	urldate = {2026-04-21},
	journal = {Chemical Engineering Progress},
	author = {ERGUN, S.},
	year = {1952},
	pages = {89},
}

@inproceedings{wen_mechanics_1966,
	title = {Mechanics of fluidization},
	volume = {62},
	url = {https://cir.nii.ac.jp/crid/1572261550060180736},
	urldate = {2026-04-21},
	booktitle = {Fluid {Particle} {Technology}, {Chem}. {Eng}. {Progress}. {Symposium} {Series}},
	author = {Wen, C. Yu},
	year = {1966},
	pages = {100--111},
}

@article{bisht_heterarchical_2024a,
	title = {Heterarchical modelling of comminution for rotary mills: part {II}—particle crushing with segregation and mixing},
	volume = {26},
	issn = {1434-7636},
	shorttitle = {Heterarchical modelling of comminution for rotary mills},
	url = {https://doi.org/10.1007/s10035-024-01450-2},
	doi = {10.1007/s10035-024-01450-2},
	language = {en},
	number = {4},
	urldate = {2026-04-20},
	journal = {Granular Matter},
	author = {Bisht, Mukesh Singh and Guillard, François and Shelley, Paul and Marks, Benjy and Einav, Itai},
	month = aug,
	year = {2024},
	pages = {87},
}

@article{bisht_heterarchical_2024,
	title = {Heterarchical modelling of comminution for rotary mills: part {I}—particle crushing along streamlines},
	volume = {26},
	issn = {1434-7636},
	shorttitle = {Heterarchical modelling of comminution for rotary mills},
	url = {https://doi.org/10.1007/s10035-024-01446-y},
	doi = {10.1007/s10035-024-01446-y},
	language = {en},
	number = {4},
	urldate = {2026-04-20},
	journal = {Granular Matter},
	author = {Bisht, Mukesh Singh and Guillard, François and Shelley, Paul and Marks, Benjy and Einav, Itai},
	month = aug,
	year = {2024},
	pages = {88},
}

@article{mordant_velocity_2000,
	title = {Velocity measurement of a settling sphere},
	volume = {18},
	issn = {1434-6036},
	url = {https://doi.org/10.1007/PL00011074},
	doi = {10.1007/PL00011074},
	language = {en},
	number = {2},
	urldate = {2026-03-30},
	journal = {The European Physical Journal B - Condensed Matter and Complex Systems},
	author = {Mordant, N. and Pinton, J.-F.},
	month = nov,
	year = {2000},
	pages = {343--352},
}

@article{preCICEv2,
	title = {{preCICE} v2: {A} sustainable and user-friendly coupling library [version 2; peer review: 2 approved]},
	volume = {2},
	url = {https://doi.org/10.12688/openreseurope.14445.2},
	doi = {10.12688/openreseurope.14445.2},
	number = {51},
	journal = {Open Research Europe},
	author = {Chourdakis, G and Davis, K and Rodenberg, B and Schulte, M and Simonis, F and Uekermann, B and Abrams, G and Bungartz, HJ and Cheung Yau, L and Desai, I and Eder, K and Hertrich, R and Lindner, F and Rusch, A and Sashko, D and Schneider, D and Totounferoush, A and Volland, D and Vollmer, P and Koseomur, OZ},
	year = {2022},
}

@article{athani_scale_2025,
  title = {Scale dependence of segregation patterns in the filling of silos},
  author = {Athani, Shivakumar and Marks, Benjy and Guillard, François and Gillespie, Alistair and Einav, Itai},
  journal = {Phys. Rev. Lett.},
  pages = {},
  year = {2026},
  month = {May},
  publisher = {American Physical Society},
  doi = {10.1103/x2mh-v8rs},
  url = {https://link.aps.org/doi/10.1103/x2mh-v8rs}
}

@article{li_hindered_2022,
	title = {Hindered settling of flocculated multi-sized particle suspension, part {I}: {Segregation} mechanism of non-flocculated particles},
	volume = {407},
	shorttitle = {Hindered settling of flocculated multi-sized particle suspension, part {I}},
	url = {https://www.sciencedirect.com/science/article/pii/S0032591022005770?casa_token=zkQ3wlGjjkQAAAAA:lyvyDkWBO8h8R_jyJLTM-3E-Dvi2NLID9qbOPPStp6Z5xgnrlK804I9CtK8dsYQpYgeweowK768},
	urldate = {2026-03-04},
	journal = {Powder Technology},
	publisher = {Elsevier},
	author = {Li, Yuan and van Zyl, Dirk},
	year = {2022},
	pages = {117683},
}

@article{athani_unifying_2024,
	title = {Unifying suspension and granular shear-induced self-diffusion},
	volume = {998},
	url = {https://www.cambridge.org/core/journals/journal-of-fluid-mechanics/article/unifying-suspension-and-granular-shearinduced-selfdiffusion/69A96BF9FB022E4D1645A924C48CF721},
	urldate = {2026-02-27},
	journal = {Journal of Fluid Mechanics},
	publisher = {Cambridge University Press},
	author = {Athani, Shivakumar and Metzger, Bloen and Mari, Romain and Forterre, Yoël and Rognon, Pierre},
	year = {2024},
	pages = {A55},
}

@article{vanderhoef_numerical_2008,
	title = {Numerical {Simulation} of {Dense} {Gas}-{Solid} {Fluidized} {Beds}: {A} {Multiscale} {Modeling} {Strategy}},
	volume = {40},
	issn = {0066-4189, 1545-4479},
	shorttitle = {Numerical {Simulation} of {Dense} {Gas}-{Solid} {Fluidized} {Beds}},
	url = {https://www.annualreviews.org/doi/10.1146/annurev.fluid.40.111406.102130},
	doi = {10.1146/annurev.fluid.40.111406.102130},
	language = {en},
	number = {1},
	urldate = {2026-02-27},
	journal = {Annual Review of Fluid Mechanics},
	author = {Van Der Hoef, M.A. and Van Sint Annaland, M. and Deen, N.G. and Kuipers, J.A.M.},
	month = jan,
	year = {2008},
	pages = {47--70},
}

@article{richardson_sedimentation_1954,
	title = {The sedimentation of a suspension of uniform spheres under conditions of viscous flow},
	volume = {3},
	url = {https://www.sciencedirect.com/science/article/pii/0009250954850159},
	number = {2},
	urldate = {2026-02-27},
	journal = {Chemical Engineering Science},
	publisher = {Elsevier},
	author = {Richardson, J. F. and Zaki, W. N.},
	year = {1954},
	pages = {65--73},
}

@article{agrawal_role_2001,
	title = {The role of meso-scale structures in rapid gas–solid flows},
	volume = {445},
	issn = {1469-7645, 0022-1120},
	url = {https://www.cambridge.org/core/journals/journal-of-fluid-mechanics/article/role-of-mesoscale-structures-in-rapid-gassolid-flows/FB50916B284346DE7B13C8F40BBA4966},
	doi = {10.1017/S0022112001005663},
	language = {en},
	urldate = {2026-02-27},
	journal = {Journal of Fluid Mechanics},
	author = {Agrawal, Kapil and Loezos, Peter N. and Syamlal, Madhava and Sundaresan, Sankaran},
	month = oct,
	year = {2001},
	pages = {151--185},
}

@book{jackson_dynamics_2000,
	title = {The dynamics of fluidized particles},
	url = {https://books.google.com/books?hl=zh-TW&lr=&id=wV9ekwf-fA8C&oi=fnd&pg=PA1&dq=The+Dynamics+of+Fluidized+Particles&ots=zDLCG2OVGF&sig=aYRkxcRgEhQbE90rkEPuAMB4sR0},
	urldate = {2026-02-27},
	publisher = {Cambridge university press},
	author = {Jackson, Roy},
	year = {2000},
}

@article{zhu_discrete_2008,
	title = {Discrete particle simulation of particulate systems: a review of major applications and findings},
	volume = {63},
	shorttitle = {Discrete particle simulation of particulate systems},
	url = {https://www.sciencedirect.com/science/article/pii/S0009250908004168?casa_token=Y_jSvKmZq7kAAAAA:lTdr58WmAO0k0ugFjETrYNCTuewl5Ac_X0-q9F3ffi-Uqq-clvb0VIU91iQMSala5twUqWBpy30},
	number = {23},
	urldate = {2026-02-27},
	journal = {Chemical Engineering Science},
	publisher = {Elsevier},
	author = {Zhu, H. P. and Zhou, Z. Y. and Yang, R. Y. and Yu, A. B.},
	year = {2008},
	pages = {5728--5770},
}

@article{jajcevic_largescale_2013,
	title = {Large-scale {CFD}–{DEM} simulations of fluidized granular systems},
	volume = {98},
	url = {https://www.sciencedirect.com/science/article/pii/S000925091300345X?casa_token=h3wEoF19UgIAAAAA:JKVdcDpyaV5VjdfI659ld4SFF__BbaxfTkzzb_zx8HCJXlSyPy6UwoNgUqfXMH1P72CXHoEheZY},
	urldate = {2026-02-27},
	journal = {Chemical Engineering Science},
	publisher = {Elsevier},
	author = {Jajcevic, Dalibor and Siegmann, Eva and Radeke, Charles and Khinast, Johannes G.},
	year = {2013},
	pages = {298--310},
}

@article{lee_twophase_2018,
	title = {A two-phase flow model for submarine granular flows: {With} an application to collapse of deeply-submerged granular columns},
	volume = {115},
	issn = {0309-1708},
	shorttitle = {A two-phase flow model for submarine granular flows},
	url = {https://www.sciencedirect.com/science/article/pii/S0309170817306528},
	doi = {10.1016/j.advwatres.2017.12.012},
	urldate = {2026-02-27},
	journal = {Advances in Water Resources},
	author = {Lee, Cheng-Hsien and Huang, Zhenhua},
	month = may,
	year = {2018},
	pages = {286--300},
}

@article{stickel_fluid_2005,
	title = {{FLUID} {MECHANICS} {AND} {RHEOLOGY} {OF} {DENSE} {SUSPENSIONS}},
	volume = {37},
	issn = {0066-4189, 1545-4479},
	url = {https://www.annualreviews.org/doi/10.1146/annurev.fluid.36.050802.122132},
	doi = {10.1146/annurev.fluid.36.050802.122132},
	language = {en},
	number = {1},
	urldate = {2026-02-27},
	journal = {Annual Review of Fluid Mechanics},
	author = {Stickel, Jonathan J. and Powell, Robert L.},
	month = jan,
	year = {2005},
	pages = {129--149},
}

@article{anderson_fluid_1967,
	title = {Fluid {Mechanical} {Description} of {Fluidized} {Beds}. {Equations} of {Motion}},
	volume = {6},
	issn = {0196-4313, 1541-4833},
	url = {https://pubs.acs.org/doi/abs/10.1021/i160024a007},
	doi = {10.1021/i160024a007},
	language = {en},
	number = {4},
	urldate = {2026-02-27},
	journal = {Industrial \& Engineering Chemistry Fundamentals},
	author = {Anderson, T. B. and Jackson, Roy},
	month = nov,
	year = {1967},
	pages = {527--539},
}

@article{marks_heterarchical_2025,
	title = {Heterarchical granular dynamics},
	volume = {27},
	issn = {1434-7636},
	url = {https://doi.org/10.1007/s10035-025-01541-8},
	doi = {10.1007/s10035-025-01541-8},
	language = {en},
	number = {3},
	urldate = {2026-02-25},
	journal = {Granular Matter},
	author = {Marks, Benjy and Athani, Shivakumar and Einav, Itai},
	month = jun,
	year = {2025},
	pages = {67},
}

@article{bisht_heterarchical_2025,
	title = {Heterarchical comminution model for {SAG} mills},
	volume = {233},
	issn = {0892-6875},
	url = {https://www.sciencedirect.com/science/article/pii/S0892687525003917},
	doi = {10.1016/j.mineng.2025.109563},
	urldate = {2026-02-25},
	journal = {Minerals Engineering},
	author = {Bisht, Mukesh Singh and Guillard, François and Shelley, Paul and Marks, Benjy and Einav, Itai},
	month = nov,
	year = {2025},
	pages = {109563},
}

@article{marks2017heterarchical,
	title = {A heterarchical multiscale model for granular materials with evolving grainsize distribution},
	volume = {19},
	number = {3},
	journal = {Granular Matter},
	publisher = {Springer},
	author = {Marks, Benjy and Einav, Itai},
	year = {2017},
	pages = {61},
}

\end{document}